\documentclass[fleqn,usenatbib]{mnras}

\usepackage{newtxtext,newtxmath}

\usepackage[T1]{fontenc}

\DeclareRobustCommand{\VAN}[3]{#2}
\let\VANthebibliography\thebibliography
\def\thebibliography{\DeclareRobustCommand{\VAN}[3]{##3}\VANthebibliography}

\usepackage{graphicx}	
\usepackage{amsmath}	

\usepackage{amssymb}	
\usepackage{longtable}
\usepackage{mathtools}

\usepackage[dvipsnames]{xcolor}

\usepackage{lscape} 

\title[74 New ZZ Cetis from {\em TESS}]{Discovery of 74 new bright ZZ Ceti stars in the first three years of {\em TESS}}

\author[A. D. Romero et al.]{Alejandra D. Romero$^{1}$\thanks{E-mail: alejandra.romero@ufrgs.br},
S. O. Kepler$^{1}$, J. J. Hermes$^{2}$, Larissa Antunes Amaral$^{1,3}$, Murat Uzundag$^{3,4}$,
\newauthor  Zs\'ofia Bogn\'ar$^{5,6}$, Keaton J. Bell$^{7,8}$, Madison VanWyngarden$^{2}$, Andy Baran$^{9,10,11}$, Ingrid Pelisoli$^{12}$,
\newauthor Gabriela da Rosa Oliveira$^1$, Detlev Koester$^{13}$, T. S. Klippel$^1$, Luciano Fraga$^{14}$, Paul A. Bradley$^{15}$, 
\newauthor Maja Vu\v{c}kovi\'{c}$^{3}$, Tyler M. Heintz$^{2}$, Joshua S. Reding$^{16}$, B. C. Kaiser$^{16}$, and St\'{e}phane Charpinet$^{17}$
\\
$^{1}$Instituto de F\'{\i}sica, Universidade Federal do Rio Grande do Sul, 91501-970 Porto Alegre, RS, Brazil\\
$^{2}$Department of Astronomy \& Institute for Astrophysical Research, Boston University, 725 Commonwealth Ave., Boston, MA 02215, USA\\
$^{3}$Instituto de F\'{\i}sica y Astronom\'{\i}a, Universidad de Valpara\'{\i}so, Gran Breta\~na 1111, Playa Ancha, Valpara\'{\i}so 2360102, Chile\\
$^{4}$European Southern Observatory, Alonso de Cordova 3107, Santiago, Chile \\
$^{5}$Konkoly Observatory, E\"otv\"os Lor\'and Research Network (ELKH), Research Centre for Astronomy and Earth Sciences, Konkoly\\ Thege Mikl\'os \'ut 15-17, H–1121, Budapest, Hungary\\
$^{6}$ MTA CSFK Lend\"ulet Near-Field Cosmology Research Group\\
$^{7}$DIRAC Institute, Department of Astronomy, University of Washington, Seattle, WA-98195, USA\\
$^{8}$NSF Astronomy and Astrophysics Postdoctoral Fellow\\
$^{9}$ARDASTELLA Research Group, Institute of Physics, Pedagogical University of Cracow, ul. Podchor\c{a}\.zych 2, 30-084 Krak\'ow, Poland\\ 
$^{10}$ Embry-Riddle Aeronautical University, Department of Physical Science, Daytona Beach, FL 32114, USA \\
$^{11}$ Department of Physics, Astronomy, and Materials Science, Missouri State University, Springfield, MO 65897, USA \\
$^{12}$Department of Physics, University of Warwick, Gibbet Hill Road, Coventry, CV4 7AL, UK\\
$^{13}$Institut f\"ur Theoretische Physik und Astrophysik, Universit\"at Kiel, D-24098 Kiel, Germany\\
$^{14}$Laborat\'orio Nacional de Astrof\'{\i}isica LNA/MCTIC, 37504-364 Itajub\'a, MG, Brazil\\
$^{15}$XCP-6, MS F-699 Los Alamos National Laboratory, Los Alamos, NM 87545\\
$^{16}$University of North Carolina at Chapel Hill, Department of Physics and Astronomy, Chapel Hill, NC 27599, USA\\
$^{17}$IRAP, Universite de Toulouse, CNRS, UPS, CNES, 14 avenue Edouard Belin, F-31400 Toulouse, France\\}

\date{Accepted XXX. Received YYY; in original form ZZZ}

\pubyear{2021}

\begin{document}
\label{firstpage}
\pagerange{\pageref{firstpage}--\pageref{lastpage}}
\maketitle

\begin{abstract}
We report the discovery of 74 new pulsating DA white dwarf stars, or ZZ~Cetis, from the data obtained by the Transiting Exoplanet Survey Satellite (TESS) mission, from Sectors 1 to 39, corresponding to the first 3 cycles. This includes objects from the Southern Hemisphere (Sectors 1--13 and 27--39) and the Northern Hemisphere (Sectors 14--26), observed with 120 s- and 20 s-cadence. Our sample likely includes 13 low-mass and one extremely low-mass white dwarf candidate, considering the mass determinations from fitting Gaia magnitudes and parallax. In addition, we present follow-up time series photometry from ground-based telescopes for 11 objects, which allowed us to detect a larger number of periods. For each object, we analysed the period spectra and performed an asteroseismological analysis, and we estimate the structure parameters of the sample, i.e., stellar mass, effective temperature and hydrogen envelope mass. We estimate a mean asteroseismological mass  of $\langle M_{\rm sis}\rangle$= 0.635$\pm$ 0.015 M$_{\odot}$, excluding the candidate low or extremely-low mass objects. This value is in agreement with the mean mass using estimates from Gaia data, which is $\langle M_{\rm phot}\rangle$= 0.631$\pm$ 0.040 M$_{\odot}$, and with the mean mass of previously known ZZ~Cetis of $\langle M_*\rangle$= 0.644 $\pm$ 0.034 M$_{\odot}$.
Our sample of 74 new bright ZZ~Cetis increases the number of known ZZ~Cetis by $\sim$20\,per\,cent.

\end{abstract}




\section{Introduction}

Variable DA white dwarf or ZZ~Ceti stars are {\bf cool} pulsating white dwarfs, with an instability strip located between effective temperatures of $\sim13\,000$~K and 10\,000 K, depending on stellar mass \citep{2017ApJS..232...23H,2017EPJWC.15201011K}. These objects show photometric variations with periods between 70 and 2000~s, and amplitudes up to 0.3 mag \citep{2008ARA&A..46..157W,2008PASP..120.1043F,2010A&ARv..18..471A,2019A&ARv..27....7C}, corresponding to spheroidal non-radial gravity modes with low harmonic degree.

Depending on their effective temperature and pulsational properties, ZZ Cetis can be classified as hot, intermediate and cool ZZ Cetis \citep{1993BaltA...2..407C,2006ApJ...640..956M}. The hot ZZ Cetis are located at the blue edge of the instability strip. They show a stable sinusoidal or sawtooth light curve, with a few modes with short periods (<350 s) and small amplitudes (1.5–20~mma). The cool ZZ Cetis, on the other hand, are located at the red edge of the instability strip, showing a collection of long periods  (up to 1500 s), with large  variation amplitudes (40–110~mma). Their light curves are non-sinusoidal and suffer from severe mode interference. Finally, the intermediate ZZ Cetis show mixed characteristics from hot and cool members. To date, there are roughly 420 ZZ Cetis known \citep[see for instance][]{2016IBVS.6184....1B,2019A&ARv..27....7C,2020AJ....160..252V,2021ApJ...912..125G}.

The excitation mechanism acting on ZZ Ceti stars is related to an opacity bump due to partial ionization of hydrogen, called the $\kappa-\gamma$-mechanism \citep{1981A&A...102..375D,1982ApJ...252L..65W}, which combines with the  convective driving mechanism \citep{1991MNRAS.251..673B,1999ApJ...511..904G} when a thick convective region develops in the outer layers. 

The Transiting Exoplanet Survey Satellite ({\em TESS}) was launched on 18 April 2018 \citep{2014SPIE.9143E..20R}, with the primary mission of searching for exoplanets around bright target stars.
Through nearly continuous stable photometry, as well as its extended sky coverage, TESS has made a significant contribution to the study of stellar pulsations in evolved compact objects  \citep[e.g.][]{2019A&A...632A..42B,2020ApJ...888...49W,2020A&A...638A..82B,2021A&A...645A.117C,2021A&A...655A..27U}, including variable hydrogen-rich DA white dwarf stars. The activities related to compact pulsators, as white dwarf and subdwarf stars, are coordinated by the TESS Asteroseismic Science Consortium (TASC), Compact Pulsators Working Group (WG8). 

In this work we present 74 new ZZ~Ceti stars discovered from the first three cycles of TESS data, from Sector 1 to Sector 39, including 120 s- and 20 s-cadence data. In addition, we perform ground-based photometry from four different telescopes for 11 objects, leading to the discovery of a new ZZ~Ceti, and in most cases increasing the number of detected pulsation periods.
This paper is organized as follows. We present the sample of 74 new ZZ Cetis, discovered from the TESS data in Section \ref{section2}. We describe the sample selection and the data reduction for the TESS data and the ground-based observations in Section \ref{section3},  including spectroscopic follow-up for 29 targets. In Section \ref{section4} we present the pulsation periods detected, and perform an asteroseismological study for our sample in Section \ref{section5}. In Section \ref{section6} we present a study of the asteroseismological properties of the 74 ZZ Cetis presented in this work, and additionally one object without TESS data. We conclude in Section \ref{conclusions} by summarizing our findings.

\section{New ZZ ceti stars}
\label{section2}

We report the discovery of 74 new bright ZZ Ceti stars from the first three years of TESS data, Sectors 1 to 39. The targets are listed in Table~\ref{tab:list1}, along with the coordinates in J2000, G magnitude, effective temperature, surface gravity, and stellar mass. The parameters are taken from various works, which used different techniques to determine the effective temperature and surface gravity or stellar mass. Also included is the object TIC~20979953 which was discovered to be variable from ground based observations (see section \ref{new-tic} for details). For those objects with more than one determination for their atmospheric parameters, we include those obtained using different techniques. 

To determine the atmospheric parameters, \citet{2008AJ....136..899S} used low-resolution spectroscopy and multi-epoch $VRI$ photometry combined with near-infrared $JHK_S$ photometry from 2MASS. They used model atmospheres from \citet{1995ApJ...443..764B}, assuming a $\log g = 8.0$, because trigonometric parallaxes were not available at the time. 

The atmospheric parameters from \citet{2009A&A...505..441K} were based on high-resolution spectra with UVES/VLT. The spectra were compared with theoretical model atmospheres from \citet{2009A&A...498..517K}.

\citet{2011ApJ...743..138G} presented the results of an spectroscopic survey of bright ($V\leq 17.5$), hydrogen-rich white dwarf stars. To derived $T_{\rm eff}$ and $\log g$ they used an updated version of the pure hydrogen model atmospheres of \citet{2005ApJS..156...47L}, that consider energy transport by convection following the MLT/$\alpha$~=~0.8 prescription of the mixing length theory \citep{2010ApJ...712.1345T} and improved Stark broadening profiles of hydrogen lines from \citet{2009ApJ...696.1755T}.

\citet{2013AJ....145..136L} performed follow-up spectroscopic observations for a sub-sample of identified white dwarf stars. They employed model atmospheres described in \citet{1995ApJ...443..764B},  with the improvements discussed in \citet{2009ApJ...696.1755T}. These are pure hydrogen, plane-parallel model atmospheres, that consider energy transport by convection following the ML2/$\alpha$~=~0.7 prescription of the mixing-length theory.

\citet{2017MNRAS.472.4173R} performed follow-up spectroscopy for a sample of white dwarfs and hot subdwarfs, extracted from an all-sky catalogue of  UV, optical and IR photometry and proper motion. For the determination of the effective temperature and surface gravity for the white dwarf stars, they employed model atmospheres from \citet{2010MmSAI..81..921K}, which adopt a MLT/$\alpha$~= ~0.8 mixing length prescription for convective atmospheres and the Stark broadening  computed by \citet{2009ApJ...696.1755T}.

Most of the data presented in Table~\ref{tab:list1} were taken from  \citet{2021MNRAS.508.3877G} and \citet{2019MNRAS.482.4570G}, where they used the Gaia DR3 and DR2 \citep{2018A&A...616A..10G} magnitudes and parallax, respectively, to determine the atmospheric parameters. They employed standard hydrogen atmosphere spectral models \citep{2011ApJ...730..128T} including the $L_{\alpha}$ red wing absorption of \citet{2006ApJ...651L.137K}. To compute the stellar mass, they used the evolutionary sequences from \citet{2001PASP..113..409F} with thick hydrogen layers and central composition C/O=50/50. 

Finally, \citet{2020ApJ...898...84K} and \citet{2020AJ....160..252V} rely on parallaxes from Gaia DR2 and photometry from the Sloan Digital Sky Survey \citep[SDSS,][]{2006ApJS..167...40E,2013ApJS..204....5K,2019MNRAS.486.2169K} and Panoramic Survey Telescope and Rapid Response System \citep[Pan-STARRS,][]{2016arXiv161205560C}. They applied the photometric technique described in \citet{1997ApJS..108..339B}, together with the pure hydrogen model atmospheres discussed in \citet{2019ApJ...876...67B} and reference therein. To derived $\log g$ and stellar mass they used white dwarf models similar to those described in \citet{2001PASP..113..409F}.

For some objects we determine the atmospheric parameter from  the spectra we obtained with the SOAR telescope (see section \ref{Murat} for details). We follow the fitting procedure described in detail in \citet{2019MNRAS.486.2169K}.

TIC~345202693 is in a binary system with a possible M main sequence star which has a large contribution in the infrared wavelengths, with a value for $G_{\mathrm{BP}} - G_{\mathrm{RP}}$ = 0.637 and an absolute magnitude of 11.83 from Gaia EDR3. Based on spectroscopic observations from the SOAR telescope,  we estimate the effective temperature of the white dwarf component. 

\begin{table*}
\centering
	\caption{List of the 74 new ZZ Ceti from TESS and TIC presented in this work. Column 1 indicates the TIC identifier. The coordinates in J2000 are in columns 3 and 4, and the G magnitude is listed in column 5. The effective temperature, $\log g$ and stellar mass determinations are listed in columns 6, 7 and 8. Data taken from the works of (1) \citep{2021MNRAS.508.3877G}, (2) \citet{2008AJ....136..899S} (3) \citet{2017MNRAS.472.4173R}, (4) \citet{2009A&A...505..441K}, (5) \citet{2020AJ....160..252V}, (6) \citet{2013AJ....145..136L}, (7) \citet{2020ApJ...898...84K}, (8) \citep{2011ApJ...743..138G}, (9)
	\citep{2019MNRAS.486.2169K}, (10)
	\citet{2019MNRAS.482.4570G} and (11) This work. The last column indicates which objects are spectroscopically confirmed DA white dwarf stars, though detection of pulsations at these temperatures implies all objects are DA white dwarfs. The object TIC~20979953 was discovered to be variable from ground-based observations}. 
	\label{tab:list1}
	\begin{tabular}{rcccccccl} 
\hline
TIC   &   RA &    DEC &  G & $T_{\rm eff}$ [K] &   $\log g$ & Mass [$M_{\odot}$]  &     Ref. & Spectrum \\
 \hline
5624184 & 09:32:48.01 & $-$37:44:28.7 & 15.95 & $11286\pm 123$ & $7.588\pm 0.018$ & 0.418 & 1 & \\ 
7675859 & 18:12:22.74 & $+$43:21:07.3  & 16.24 & $12240\pm 214$ & $8.479\pm 0.023$ & 0.909  & 1 &  \\ 
8445665 & 16:24:36.81 & $+$32:12:52.8 & 16.72 & $11385\pm 235$ & $7.947\pm 0.040$ & 0.574  &  1  &  DA\\ 
13566624 & 08:51:34.85 & $-$07:28:28.3 & 16.44  & $13634\pm 298$ & $8.188\pm 0.029$ & 0.724 & 1 &  \\
20979953 & 15:33:32.96 & $-$02:06:55.7 & 16.53 & $11859\pm 236$ & $7.969\pm 0.039$ & 0.587 & 1  &  \\
 $\cdots$  & $\cdots$  & $\cdots$   & $\cdots$  & $11212\pm 54$ & $7.89\pm 0.01$ & 0.540 & 7  &  \\
 $\cdots$  & $\cdots$  & $\cdots$   & $\cdots$  & $1132\pm 250$  & $8.190\pm 0.022$ & 0.713 & 9  & DA \\
21187072 & 18:26:06.04 & $+$48:29:11.3 & 16.28 & $11808\pm 228$  & $7.235\pm 0.025$ &  0.314 & 10  & DA \\ 
24603397 & 05:22:40.66 & $-$08:02:29.7 & 14.71 & $11832\pm 136$ & $7.842\pm 0.017$ & 0.517 & 1  & DA\\ 
29862344 & 01:37:15.16 & $-$17:27:22.7 & 15.25 & $11613\pm 192$ & $8.14\pm 0.06 $  & 0.682 &  2 & DA  \\
33717565 & 04:05:36.39 & $-$76:28:28.1 & 16.52 & $10675\pm 172$ & $7.639\pm 0.031$ & 0.433  & 1 & DA \\
46847635 & 09:29:16.70 & $-$08:40:32.2 & 16.75 & $12018\pm 344$ & $7.979\pm 0.048$ & 0.593 & 1  &  \\
55650407 & 04:55:27.27 & $-$62:58:44.6 & 14.99 & $11838\pm 150$ & $7.945\pm 0.019$ & 0.574 & 1 &  \\
$\cdots$ & $\cdots$ & $\cdots$ &  $\cdots$ & $12134\pm 67$ & $7.906\pm 0.002$ & 0.556 & 11  &  DA\\
63281499 & 22:28:58.15 & $-$31:05:53.7 & 15.61 & $11712\pm 166$ & $7.981\pm 0.025 $ & 0.594 & 1 &  \\ 
$\cdots$   & $\cdots$  & $\cdots$  & $\cdots$ & $12200\pm 200$ & $8.02\pm 0.06 $ & 0.616 & 3 & DA \\ 
65144290 & 07:11:14.04 & $-$25:18:15.0 & 14.47 & $11208\pm 180$ & $8.12\pm 0.060$ &  0.670 & 10 &  DA \\ 
72637474 & 02:08:07.86 & $-$29:31:38.0 & 15.92 & $10214\pm 112$ & $7.209\pm 0.025$ & 0.297 & 1  &  \\
  $\cdots$ & $\cdots$ &  $\cdots$    & $\cdots$  & $11769\pm 250$ & $7.54\pm 0.038$ & 0.413 & 4  & DA \\ 
79353860 & 21:18:15.52 & $-$53:13:22.7 & 15.92 & $11284\pm 196$ & $7.970\pm 0.032$ & 0.587 & 1 &  \\ 
$\cdots$ & $\cdots$ &  $\cdots$    & $\cdots$  &$11372\pm 44$ & $7.982\pm 0.004$ & 0.548 & 11 & DA \\ 
116373308 & 03:02:11.43 & $+$48:00:13.6 & 16.33 & $12057\pm 265$ & $8.043\pm 0.033$ & 0.631 & 1 & DA \\ 
 $\cdots$  &  $\cdots$ &  $\cdots$    &  $\cdots$ & $11551\pm 60$  & $\cdots$  & 0.614 & 5  &  \\  
141976247 & 06:25:27.47 & $-$75:40:41.7 & 15.58 & $13121\pm 241$ &  $8.204\pm 0.018$ &  0.733 & 1 & DA \\ 
149863849 & 17:43:49.28 & $-$39:08:25.9 & 13.53 & $11604\pm  206$ & $8.087\pm 0.027$ &  0.657 & 1  & DA \\ 
156064657 & 00:37:23.75 & $-$48:21:55.9 & 16.60 & $10193\pm  131$ & $7.295\pm 0.034$ & 0.320  & 1 & DA \\ 
158068117 & 06:00:52.91 & $-$46:30:41.1 & 16.09 & $12719\pm 210$ & $7.434\pm 0.026$ &  0.376 & 1  &  \\ 
167486543 & 04:48:32.11 & $-$10:53:49.9 & 16.23 & $12187\pm 252$ & $8.547\pm 0.028$ &  0.953 & 1 & DA \\ 
188087204 & 10:46:27.80 & $-$25:12:15.8 & 16.83 & $10052\pm 218$ & $7.583\pm 0.055$ & 0.412 & 1  &  \\ 
207206751 & 03:13:18.66 & $-$56:07:35.0 & 14.62 & $10968\pm 250$ & $7.996\pm 0.038$ & 0.601 & 10  & DA\\ 
220555122 & 02:56:21.34 & $-$63:28:40.2 & 15.87 & $11827\pm 350$ & $8.169\pm 0.045$  & 0.708 & 1  & DA \\ 
229581336 & 18:01:15.37  & $+$72:18:49.0  & 16.05 & $14634\pm 403$ & $7.425\pm 0.035$ & 0.382 &   1 & DA \\ 
$\cdots$   & $\cdots$  & $\cdots$  & $\cdots$ & $11075\pm 32$ & $8.082\pm 0.022$ & 0.648  & 11 & DA \\
230029140 & 19:28:53.87  & $+$61:05:48.7  & 16.45 & $11655\pm 230$ & $7.987\pm 0.051$ & 0.597 & 1 & DA \\ 
230384389 & 19:03:19.56 & $+$60:35:52.6 & 15.04 & $11366\pm  89$ & $8.09\pm 0.014$ &  0.658  & 10 &  DA\\ 
231277791 & 02:49:18.23 & $-$53:34:35.4 & 16.46 & $11623\pm 235$ & $8.080 \pm 0.035$ & 0.653  & 1 & DA  \\ 
232979174 & 14:34:17.88 & $+$65:39:59.5 & 16.14 & $12776\pm 234$ & $7.749\pm 0.025$ & 0.473 & 1 & DA \\ 
238815671 & 21:52:11.62 & $-$63:32:36.4 & 16.12 & $11693\pm 250$ & $7.944\pm 0.038$ & 0.573 & 10 &  DA \\ 
261400271 & 06:51:01.30 & $-$80:34:09.6 & 14.90  & $13670\pm 250$ & $8.399\pm 0.038$ & 0.859 & 10  &  DA\\ 
273206673 & 04:33:50.99 & $+$48:50:39.2 & 15.35 & $11433\pm 221$ & $7.966\pm 0.035$ & 0.585 & 1 &  \\ 
282783760 & 13:14:26.82 & $+$17:32:09.2  & 16.30 & $12111\pm 268$ & $8.026\pm 0.032$ & 0.622 & 1 & DA \\ 
287926830 & 21:50:40.62 & $+$30:35:34.1 & 15.94 & $11429\pm 79$ & $\cdots$ & 0.562 & 5 & \\
304024058 & 09:22:56.24 & $-$68:16:48.8 & 16.10 & $11368\pm 208$ & $7.955\pm  0.033$ & 0.578 & 1  & DA \\ 
313109945 & 14:05:40.57 & $+$74:38:59.3  & 15.59 &  $9059\pm 134$ & $7.49\pm 0.04 $  &  0.380 &  1 & DA \\ 
317153172 & 23:22:32.11 & $-$83:13:14.2 & 16.47 & $11813\pm 314$ & $8.032\pm 0.042$ &  0.624 & 1 &  DA\\ 
317620456 & 19:21:82.42& $+$27:40:25.4 & 15.04 & $10566\pm 67$  & $\cdots$    & 0.603 & 7  &  \\
 $\cdots$  &  $\cdots$ &  $\cdots$    &  $\cdots$ & $11060\pm 163$ & $8.10\pm 0.05$   & 0.660    & 6 & DA \\
343296348 & 17:43:44.00 & $-$74:24:37.5 & 15.85 & $11597\pm 150$ & $7.968\pm 0.023$ & 0.586 & 1 &  DA \\ 
344130696 & 18:37:08.31 & $-$76:59:05.9 & 15.39 & $10829\pm 116$ & $7.177\pm 0.018$ &  0.293 & 1  & DA \\ 
345202693 & 18:48:28.03 & $-$74:27:60.0 & 16.56 & $11332\pm 250$ &  $\cdots$  & $\cdots$   & 1 &  DA+IR \\
353727306 & 02:40:29.66 & $+$66:36:37.1 & 15.60 & $11874\pm 250$ & $8.019\pm 0.038$ & 0.617 & 10 & DA \\ 
370239521 & 21:50:24.19 & $-$53:58:37.2 & 14.65 &  $11191\pm 978$ & $8.23\pm 0.44$ & 0.733 & 11  & DA+M \\ 
380298520 & 20:13:43.26 & $+$34:13:56.0 & 15.68 & $11834\pm 170$ & $8.405\pm 0.021$ & 0.861    & 1  & DA \\
 $\cdots$  &  $\cdots$ &  $\cdots$    &  $\cdots$ & $11440\pm 118$ & $\cdots$   & 0.854 & 5  &  \\
394015496 & 21:58:23.88 & $-$58:53:53.8 & 15.81 & $11639\pm 141$ & $8.004\pm 0.022$  & 0.607 & 1 & DA \\
415337224 & 03:54:54.26 & $+$07:46:06.3 & 16.54 & $16380\pm 308$ & $7.87\pm 0.06$ & 0.55 & 8  &  \\
$\cdots$  &  $\cdots$ &  $\cdots$  & $\cdots$ & $10600\pm 300$ & $\cdots$   & 0.563 & 10  & DA \\   
428670887 & 11:58:40.65 & $-$20:29:51.2 & 16.01 & $11826\pm 175$ & $8.062\pm 0.026$ & 0.642 & 1  & DA \\ 
\hline
\end{tabular}
\end{table*}

\begin{table*}
\centering
\contcaption{}
	\begin{tabular}{rcccccccl} 
\hline
TIC   &   RA &    DEC &  G & $T_{\rm eff}$ [K] &   $\log g$ & Mass [$M_{\odot}$]  &     Ref. &  Spectrum \\
\hline
441500792 & 03:06:48.35 & $-$17:23:32:9 & 16.68 & $11393\pm 273$ & $8.046\pm 0.046$ &  0.632 & 1 & \\
442962289 & 05:25:47.64 & $-$17:33:49.9 & 16.51 & $11945\pm 252$ &  $8.416\pm 0.031$ & 0.868 & 1 & \\
610337553 & 00:55:46.72 & $-$15:04:52.7 & 17.36 &  $10680\pm 380$ & $7.854\pm 0.089$ & 0.520 & 1  &  \\
631161222 & 01:26:24.73 & $-$71:17:12.0 & 16.96 & $11435\pm 362$ & $7.934\pm 0.059$ & 0.566 & 1  &  \\ 
631344957 & 02:13:28.27 & $-$64:37:08.9 & 16.98 & $11574\pm 311$ & $7.995\pm 0.050$ &  0.602 & 1  &   \\ 
632543879 & 02:28:23.39 & $+$13:47:27.3 & 16.98 & $11850\pm 380$ & $8.036\pm 0.055$ & 0.626 & 1 &  \\
651462582 & 03:07:33.09 & $-$46:53:16.3 & 17.08 & $11210\pm 340$ & $7.930\pm 0.064$ & 0.564 & 1  &  \\ 
661119673 & 04:42:58.31 & $+$32:37:15.6 & 17.37 & $10668\pm 345$ & $7.881\pm 0.082$ & 0.535 & 1 &  \\ 
685410570 & 05:00:11.50 & $-$50:46:12.4 & 17.04 & $11257\pm 335$ & $7.944\pm 0.059$ & 0.572 & 1  &  \\
686044219 & 04:21:48.96 & $-$35:58:49.8 & 17.13 & $11477\pm 356$  & $8.019\pm 0.058$  & 0.615 & 1 &   \\
712406809 & 06:39:17.24 & $+$01:13:29.5 & 16.22 & $10669 \pm 298$ & $7.914\pm 0.065$ & 0.553 & 7 & DA \\ 
724128806 & 05:37:24.22 & $-$80:45:49.7 & 17.48 & $9776\pm 239$ &  $7.595\pm  0.065$ &  0.415 & 1  &  \\
733030384 & 05:32:03.91 & $-$65:36:09.9 & 16.89 & $12059\pm 193$ & $8.001\pm 0.028$ &  0.606 & 1 &  \\
800153845 & 08:55:07.25 & $+$06:35:40.0 & 16.656 & $10501\pm 273$ & $7.902\pm 0.070$ & 0.546 & 1  &         \\
$\cdots$  &  $\cdots$ &  $\cdots$  & $\cdots$     & $11119\pm 46$ & $8.365\pm 0.030$ & 0.820 & 9 & DA \\ 
804835539 & 08:54:57.51 & $-$76:46:21.9 & 16.906 & $15296\pm 476$ &  $8.078\pm 0.046$ & 0.659 & 1  &  \\
804899734 & 08:32:58.10 & $-$76:01:05.9 & 17.40 & $11585\pm 305$ & $ 7.988\pm 0.053$ & 0.598 & 1  &  \\
951016050 & 12:14:11.95 & $-$34:58:45.9 & 17.03 & $11292\pm 263$ & $7.974\pm 0.051$ &  0.589 & 1  &  \\
1001545355 & 14:13:53.96 & $+$71:36:12.6 & 16.99 & $11244\pm 232$ & $7.653\pm 0.036$ &  0.439 & 1  &  \\
1102242692 & 15:28:09.16 & $+$55:39:16.1 & 17.09 &  $10343\pm 223$ & $7.459\pm 0.051$ & 0.372 & 1 &  \\     
$\cdots$  &  $\cdots$ &  $\cdots$  & $\cdots$  & $11180\pm 184$ & $7.86\pm 0.07$  &  0.530 & 8  & DA \\
1102346472 & 14:53:23.52 & $+$59:50:56.2 & 17.16 & $12102\pm 345$ & $8.067\pm 0.045$ & 0.646 & 1  &  \\
$\cdots$  &  $\cdots$ &  $\cdots$  & $\cdots$ &  $11217\pm 250$  & $7.97\pm 0.038$ & 0.589 & 9  & DA \\
1108505075 & 15:44:55.68 & $-$69:09:10.4 & 16.993 & $11369\pm 225$ & $7.918\pm 0.043$ &  0.557 & 1  &  \\ 
1173423962 & 14:41:14.41 & $-$38:46:29.7 & 17.38 & $10719\pm 414$ & $7.920\pm  0.093$ &  0.556 & 1 &  \\
1201194272 & 16:33:58.75 & $+$59:12:06.6 & 17.13 & $11705\pm 416$ & $8.018\pm 0.063$ & 0.616  & 1  &  \\
1309155088 & 16:54:26.50 & $+$23:52:41.5 & 16.99 & $11487\pm 273$ & $8.109\pm 0.045$ &  0.670 & 1  &  \\
$\cdots$  &  $\cdots$ &  $\cdots$  & $\cdots$ & $10700\pm 91$ & $8.033\pm 0.013$  &  0.622  & 7  &  DA\\
1989258883 & 20:14:39.52 & $-$56:55:01.7 & 16.62 & $11172\pm 188$ & $7.983\pm 0.037$ & 0.594    & 1 & DA \\ 
1989866634 & 20:43:11.73 & $-$46:10:48.5 & 17.38 & $11503\pm 419$ & $7.996\pm 0.075$ & 0.602    & 1 & DA \\
2026445610 & 21:24:21.14 & $-$63:10:12.4 & 17.27 & $11737\pm 274$ & $8.037\pm 0.046$ &  0.627 & 1 &  \\
2055504010 & 22:45:54.79 & $-$45:00:58.9 & 16.84 & $11050\pm 239$ & $7.900\pm 0.049$ & 0.546  & 1 &  \\
\hline
\end{tabular}
\end{table*}

\begin{figure*}
	\includegraphics[width=0.9\textwidth]{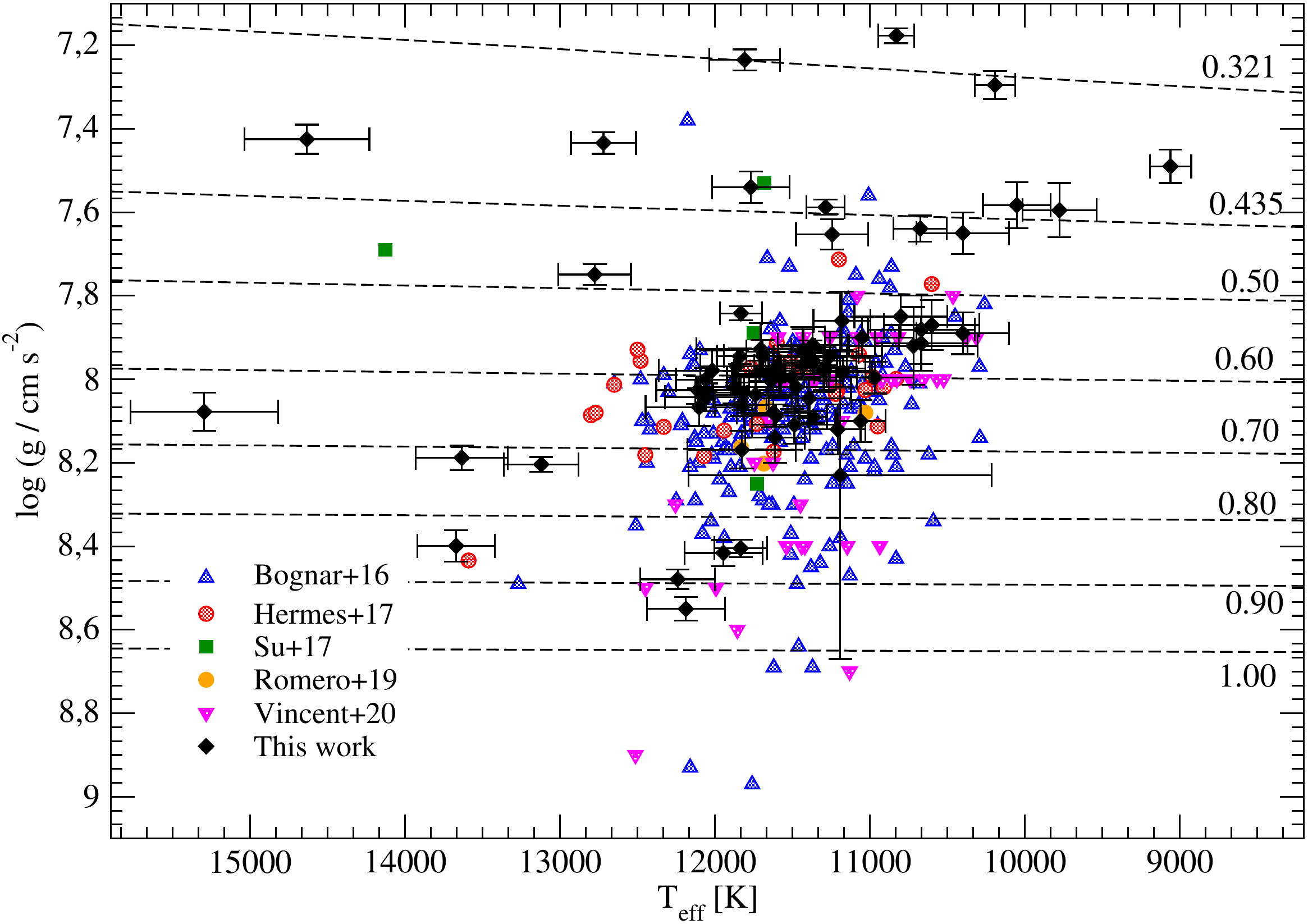}
    \caption{Distribution of ZZ Ceti stars on the $T_{\rm  eff}-\log g$ plane. The coloured symbols correspond to known ZZ Ceti stars, taken from \citet{2016IBVS.6184....1B} (blue up-triangle), \citet{2017ApJS..232...23H} (red circle), \citet{2017ApJ...847...34S} (green square), \citet{2019MNRAS.490.1803R} (orange circle) and \citet{2020AJ....160..252V} (magenta down-triangle). The objects observed in this work are depicted with black circles. We include evolutionary tracks (dashed lines) with stellar masses between 0.435~M$_{\odot}$ and 0.9~M$_{\odot}$ from \citet{2019MNRAS.484.2711R} and 0.321~M$_{\odot}$ from \citet{2014A&A...571A..45I}.} 
    \label{ZZCetis}
\end{figure*}

The location of the 74 new ZZ Ceti stars in the $T_{\rm eff}-\log g$ plane is presented in Figure \ref{ZZCetis}. For the targets with more than one determination for the atmospheric parameters in Table~\ref{tab:list1}, we adopt that obtained from photometry and parallax from Gaia for Figure~\ref{ZZCetis}.
The sample of ZZ Ceti stars known to date is depicted in this figure, and was extracted from the works of \citet{2016IBVS.6184....1B, 2017ApJS..232...23H, 2017ApJ...847...34S, 2019MNRAS.490.1803R, 2020AJ....160..252V}. The values for effective
temperature and surface gravity derived from spectroscopy were corrected by 3D convection 
\citep{tremblay} for all objects \citep{2019A&ARv..27....7C}. 
Most of the objects from our sample lay around the 0.6~M$_{\odot}$ track, showing canonical masses. Note that there are 13 objects with stellar masses in the range of 0.30~$\leq$~M/M$_{\odot} \leq$~0.45, which correspond to low-mass white dwarfs \citep{Kilic2007,2016A&A...595A..35I,Pelisoli2019} and can harbour either a He/C/O- or a He-core, depending on the evolution of the progenitor star \citep{2021MNRAS.tmp.3166R}. 
TIC~345202693 has a photometric stellar mass below 0.3~M$_{\odot}$, and is a possible extremely low-mass (ELM) white dwarf variable. Finally, there are three objects with masses above 0.8~M$_{\odot}$; TIC~7675859 is the most massive object of our sample, with a photometrically determined mass of 0.909~M$_{\odot}$.

\section{Observations and Data reduction}
\label{section3}

We selected the targets for the sample of DA white dwarf stars from \citet{2019MNRAS.482.4570G} with G~$\leq$~17.5\,mag that were targeted by the TESS satellite in Sectors 1--39, with 120~s and/or 20~s-cadence. This includes both the southern (Sectors 1--13) and the northern (Sectors 14--26) hemispheres, with 120~s cadence. From Sector 27 onward, the satellite turned back to the Southern Hemisphere, and data with 20~s cadence became available for a subset of objects. In addition, we performed photometric observations from ground based telescopes with a cadence smaller than 45~s for a small subset (11) to confirm variability and to look for new periodicities. 
Finally, we performed spectroscopic observations for another subset (29). These data were used to improve the determination of the atmospheric parameters for some targets, and in all cases to confirm that our targets are spectroscopically confirmed DA white dwarfs. A detail description of the observations and data analysis is presented in the sections below.  

\subsection{TESS data}


We downloaded all 2-min- and 20-s-cadence light curves of over 8\,300 known white dwarfs and white dwarf candidates \citep{2019MNRAS.482.4570G,2021MNRAS.508.3877G} brighter than G~$\leq$~17.5\,mag from The Mikulski Archive for Space Telescopes, which is hosted by the
Space Telescope Science Institute (STScI)\footnote{http://archive.stsci.edu/} in FITS format. The
data were processed based on the
Pre-Search Data Conditioning pipeline \citep{2016SPIE.9913E..3EJ}. We
extracted times and fluxes (PDCSAP FLUX) from the FITS
files. The times are given in barycentric corrected dynamical
Julian days \citep[BJD – 2457000, corrected for leap seconds, see][]{2010PASP..122..935E}. The fluxes were converted into fractional variations from the mean, that is, differential flux $\Delta I/I$, and transformed into amplitudes in parts-per-thousand (ppt). The ppt unit corresponds to the milli-modulation amplitude (mma) unit\footnote{1 mma= 1/1.086 mmag= 0.1\% = 1 ppt; see, e.g., \citet{2016IBVS.6184....1B}.}. We sigma-clipped the data at 5$\sigma$ to remove the outliers that appear above five times the median of intensities, that is, that depart from the median by 5$\sigma$. 

We calculated their Fourier
transforms (FTs) and examined them for pulsations or binary
signatures above the 1/1000 false-alarm probability (FAP), calculated reshuffling the data 1000 times, but maintaining the same time base, and calculating their Fourier transform, selecting the highest peak. For pre-whitening, we employed our customized
tool, in which, using a 
nonlinear least-squares (NLLS) method,
we simultaneously fit each pulsation frequency in a waveform
$A_i\sin(\omega_1 t + \phi)$, with $\omega = 2\pi / P$, and $P$ the period. This iterative process was run starting with the highest peak until no peak appeared above the 0.1\% false-alarm probability significance threshold. We analysed the concatenated light curve from different sectors, if observed. The FAP was again calculated by randomizing the observations, that is, shuffling the observations one thousand times and recalculating the FTs. We calculated the amplitude at which there was a 0.1\%= 1/1000 probability of any peak being due to noise \citep[e.g.][]{1993BaltA...2..515K}. 

Because of the large pixel scale of TESS, the flux corresponding to the white dwarf ranged from CROWDSAP=0.021--0.985, meaning the total flux from the white dwarf in the extracted aperture ranged from 2.1\,per\,cent to 98.5\,per\,cent. To confirm the variations are from the white dwarf, we checked all stars around 120"x120" in Gaia EDR3 for other possible variables or parallax and proper motion companions. In the rare cases where a white or blue star was found, we searched for variability in every pixel of the aperture, as these might show variability on similar timescales. None was found. All PDCSAP flux values are corrected for the crowding via the CROWDSAP value, so the reported amplitudes have been corrected for flux dilution.

The third year of the TESS mission started with Sector~27. From this sector on, besides the 2-min-cadence data, some objects were observed with 20-s cadence, increasing the frequency resolution in the Fourier transform. We analyze 20-s data for {\bf 15} out of {\bf 74} ZZ Cetis in this sample.


\subsection{Ground-based photometric observations}

\begin{table}
	\centering
	\caption{Journal of photometric observations from ground-based facilities. We list the target, telescope, date of observation, exposure time and total observation time, in columns 1, 2, 3, 4 and 5, respectively. The sizes of the telescopes are 1-m, 4.1-m, 1.83-m and 1.6-m, for the Konkoly observatory, SOAR, Perkins and OPD, respectively.}
	\label{tab:GB}
	\begin{tabular}{rcccc} 
\hline
TIC & Telescope & Run stars (UT) & t$_{\rm exp}$ (s) & $\Delta$t (h) \\
\hline 
7675859 & Konkoly & 2020-08-20 & 30 & 5.45\\
           &         & 2020-08-21 & 30 & 5.95\\
           &         & 2020-08-22 & 30 & 4.28\\
           &         & 2020-08-23 & 45 & 3.96\\
           &         & 2020-08-25 & 45 & 5.89\\
           &         & 2020-08-26 & 45 & 6.14\\
20979953 & OPD & 2020-06-14 & 17 & 3.4 \\
           &      & 2020-11-27 & 10 & 3.36 \\           
55650407 & SOAR & 2020-11-26 & 10 & 3.72 \\
232979174 & Konkoly & 2021-07-05 & 45 & 5.05 \\
           &         & 2021-07-06 & 45 & 5.38 \\
           &         & 2021-07-07 & 30 & 4.81 \\
           & Perkins & 2021-08-07 & 10 & 1.78 \\
273206673 & Konkoly & 2020-09-11 & 30 & 4.68\\
           &         & 2020-10-08 & 30 & 6.88\\
           &         & 2020-12-11 & 45 & 4.63\\
           &         & 2020-12-13 & 30 & 4.23\\
           &         & 2020-12-14 & 30 & 3.50\\
           &         & 2020-12-14 & 45 & 2.53\\
282783760 & OPD & 2021-06-14 & 17 & 3.1 \\           
304024058 & SOAR & 2020-12-02 &  10  & 3.46    \\   
313109945 & Konkoly & 2020-06-11 & 30 & 2.32\\
           &         & 2020-06-13 & 30 & 4.71\\
           &         & 2020-07-04 & 30 & 4.49\\
           &         & 2020-07-05 & 30 & 5.18\\
           &         & 2020-07-07 & 30 & 4.87\\
370239521 & OPD & 2020-06-14 & 17 & 3.4 \\
           &      & 2020-11-27 & 10 & 3.36 \\
1989866634 & OPD & 2021-05-09 & 40 & 3.1 \\
           &     & 2021-06-12 & 40 & 3.38  \\
2055504010 & OPD & 2021-06-13 & 20 & 2.6 \\
           &     & 2021-06-14 & 20 & 3.2 \\
\hline 
\end{tabular}
\end{table}

Ground-based follow-up photometry was performed for 11 objects with four different telescopes: the 1~m at Konkoly Observatory in Hungary, the 1.6~m Pekin-Elmer telescope at the Pico do Dias Observatory in Brazil, the 1.83-m Perkins telescope in the United States, and the 4.1~m SOAR telescope in Chile. The journal of time-series photometric observations is presented in Table~\ref{tab:GB}. 


For four objects, we performed observations with the 1~m Ritchey--Chr\'etien--Coud\'e telescope located at the Piszk\'estet\H o mountain station of Konkoly Observatory, Hungary. We obtained data with a Spectral Instruments 1100S CCD camera in white light. The exposure times were selected to be either 30 or 45\,s. We reduced the raw data frames the standard way utilizing \textsc{IRAF} tasks: we performed bias and flat field corrections before the aperture photometry of field stars. We fitted low-order polynomials to the resulting light curves, correcting for long-period instrumental and atmospheric trends, and finally, we converted the observational times of every data point to barycentric Julian dates in barycentric dynamical time (BJD$_\mathrm{{TDB}}$) using the applet of \citet{2010PASP..122..935E}\footnote{\url{http://astroutils.astronomy.ohio-state.edu/time/utc2bjd.html}}.


In addition, we employed Goodman image mode on the 4.1-m Southern Astrophysical Research (SOAR) Telescope in Chile. We used read out mode 200 Hz ATTN2 with the CCD binned $2\times2$, with a ROI reduced to 800$\times$800. All observations were obtained with a red blocking filter S8612. The integration times varied from 10 to 15\,s, depending on the magnitude of the object and the weather conditions. Note that with this configuration, the read-out time is $\sim5$\,s.

Five objects were observed using the IxON camera on the 1.6-m Perkin Elmer Telescope at the Pico dos Dias Observatory, in Brazil. We used a red blocking filter BG40. The integration times varies from 20 to 45~s, depending on the magnitude of the object, with a read-out time of less than 1 s.

For one object we obtained a light curve using the Perkins Re-Imaging SysteM (PRISM) mounted on the 1.8m Perkins Telescope Observatory (PTO) on Anderson Mesa outside of Flagstaff, Arizona. We used a red-cutoff {\em BG40} filter with 10-s exposures, minimizing readouts by windowing the CCD to 410 $\times$ 380 pixels.

We reduced the data with the software \textsc{IRAF}, and perform  aperture photometry with DAOFOT. We extracted light curves of all bright stars that were observed simultaneously in the field. Then, we divided the light curve of the target star by the light curves of all comparison stars to minimize effects of sky and transparency fluctuations. To look for periodicities in the light curves, we calculate the Fourier transform (FT) using the software \texttt{PERIOD04} \citep{2004IAUS..224..786L}. We accepted a frequency peak as significant
if its amplitude exceeds the 0.1\% FAP. We
then use the process of pre-whitening the light curve by subtracting out of the data a sinusoid with the same frequency, amplitude, and phase of the highest peak, and then computing the FT for the residuals. We redo this process until we have no new significant signals.


\subsection{Ground-based spectroscopic observations}
\label{Murat}

To confirm they are DA white dwarfs and to improve the determinations of the atmospheric parameters, we obtained follow-up spectroscopic observations for 29 objects from our new ZZ Ceti stars sample. Many of these observations were organized through Working Group 8 on compact objects of the TESS Asteroseismic Consortium\footnote{\url{https://tasoc.dk/}} and are detailed in Table~\ref{tablespectroscopy}.

\begin{table}
\setlength{\tabcolsep}{2pt}
\renewcommand{\arraystretch}{1.2}
\centering
\caption{Log of spectroscopic observations.}
\begin{tabular}{rcccc}
\hline
TIC & UT Date & Grating & Exp. &  Telescope/Inst. \\
    &         & (l mm$^{-1}$) & (sec) &   \\
\hline   
21187072 & 2021-03-23 & 300 & 1200 & LDT/DEVENY \\   
24603397 & 2021-09-21 & 930 & 900 & SOAR/GOODMAN    \\
29862344 & 2019-06-17 & 930 & 1500 & SOAR/GOODMAN \\
$\cdots$ & 2021-10-11 & 930 & 900 & SOAR/GOODMAN \\
33717565 & 2021-09-21 & 930 & 3000 & SOAR/GOODMAN    \\
55650407 & 2019-12-05 & 930 & 1620 & SOAR/GOODMAN    \\
$\cdots$ & 2020-12-07 & 400 & 720 & SOAR/GOODMAN    \\
63281499 & 2019-08-22 & 600 & 600 & DUPONT/B\&C     \\
$\cdots$ & 2019-12-05 & 930 & 1080 & SOAR/GOODMAN    \\
65144290 & 2021-03-05 & 400 & 400 & SOAR/GOODMAN    \\
$\cdots$ & 2021-09-21 & 930 & 600 & SOAR/GOODMAN    \\
79353860 & 2018-06-02 & 930 & 1080 & SOAR/GOODMAN    \\
$\cdots$ & 2021-09-21 & 400 & 1800 & SOAR/GOODMAN    \\
116373308 & 2020-10-19 & 300 & 840 & LDT/DEVENY \\
149863849 & 2021-09-21 & 930 & 900 &  SOAR/GOODMAN    \\
156064657 & 2021-09-21 & 930 & 2400 & SOAR/GOODMAN    \\
167486543 & 2021-10-12 & 400 & 540 &  SOAR/GOODMAN \\
207206751 & 2021-03-05 & 400 & 900  & SOAR/GOODMAN    \\
$\cdots$  & 2021-09-21 & 930 & 1200 & SOAR/GOODMAN    \\
220555122 & 2021-03-05 & 400 & 1200 &  SOAR/GOODMAN    \\
$\cdots$  & 2021-09-21 & 930 & 2400 & SOAR/GOODMAN    \\
229581336 & 2021-03-23 & 300 & 1920 & LDT/DEVENY \\
230029140 & 2021-03-23 & 300 & 540 & LDT/DEVENY \\
231277791 & 2021-09-21 & 930 &  2400 & SOAR/GOODMAN    \\
232979174 & 2021-03-23 & 300 & 900 & LDT/DEVENY \\
238815671 & 2019-06-17 & 930 & 3000 & SOAR/GOODMAN    \\  
$\cdots$  & 2021-06-19 & 400  & 500  & SOAR/GOODMAN \\
261400271 & 2021-03-05 & 400 & 500  &  SOAR/GOODMAN    \\
304024058 & 2019-12-05 & 930  & 1440  & SOAR/GOODMAN \\
343296348 & 2019-08-18 & 930  & 720   & SOAR/GOODMAN \\
344130696 & 2019-08-18 & 930  & 1440  & SOAR/GOODMAN \\
$\cdots$  & 2021-10-11 & 930 & 900 &  SOAR/GOODMAN \\
353727306 & 2020-09-22 & 300 & 1500 & LDT/DEVENY \\
370239521 & 2019-06-17 & 930  & 540  & SOAR/GOODMAN \\
$\cdots$  & 2021-06-19 & 400  & 350  & SOAR/GOODMAN \\
380298520 & 2020-09-22 & 300 & 1200 & LDT/DEVENY \\
394015496 & 2018-08-31 & 930  & 900 & SOAR/GOODMAN \\
$\cdots$ & 2021-06-19 & 400  &  500 & SOAR/GOODMAN \\
428670887 & 2021-06-18 & 400 & 1200  & SOAR/GOODMAN \\
1989258883 & 2021-06-19 & 400 &  900 &  SOAR/GOODMAN \\
\hline 
\label{tablespectroscopy}
\end{tabular}
\end{table}

One ZZ Ceti, TIC~63281499, was observed with the Boller and Chivens (B\&C) spectrograph mounted at the 2.5-meter (100-inch) Ir\'ene du Pont telescope at Las Campanas Observatory in Chile\footnote{For a description of instrumentation, see: \url{http://www.lco.cl/?epkb_post_type_1=boller-and-chivens-specs}}.
The B\&C spectra were obtained using the 600 lines/mm grating corresponding to the central wavelength of 5000~\AA, and covering a wavelength range from 3427 to 6573~\AA. We used a 1~arcsec slit, which provided a resolution of 3.1~\AA. 
The data from Dupont@B\&C was reduced and analysed using \textit{PyRAF}\footnote{\url{http://www.stsci.edu/institute/software_hardware/pyraf}} \citep{pyraf2012} procedures with the following way: 
First, bias correction and flat-field correction have been applied. 
Then, the pixel-to-pixel sensitivity variations were removed by dividing each pixel with the response function. 
After this reduction was completed, we have applied wavelength calibrations using the frames obtained with the internal HeAr comparison lamp. 
In a last step, flux calibrations were applied using the standard star EG\,274. The signal-to-noise ratio (SNR) of the final spectra is around 65 (see Table~\ref{tablespectroscopy}).

Additionally, seven new ZZ Cetis were observed with the DeVeny Spectrograph mounted on the 4.3-m Lowell Discovery Telescope \citep[DT][]{2014SPIE.9147E..2NB} in Happy Hack, Arizona, United States. Using a 300 line 1/mm grating we obtain a roughly 4.5 resolution. Our spectra were debiased and flat-fielded using standard STARLINK routines \citep{2014ASPC..485..391C}, were optimally extracted \citep{1986PASP...98..609H} using the software PAMELA, and were wavelength-calibration (including a heliocentric correction) using MOLLY \citep{1989PASP..101.1032M}. 

The majority of our southern spectroscopic observations have been obtained using the Southern Astrophysical Research (SOAR) Telescope and the Goodman spectrograph
\citep{clemens2004}, situated at Cerro Pach\'on, Chile. 
We use two main setups: in our lower-resolution setup, we use the 400\,l/mm grating with the blaze wavelength 5500 \AA\ (M1: 3000-7050 \AA) with a slit of 1 arcsec. This setup provides a resolution of about 5~\AA. Most commonly, we used the 930\,l/mm grating (M2: 3850-5550 \AA) with a slit of 0.46 arcsec. This setup provides a resolution of about 2~\AA. Table~\ref{tablespectroscopy} outlines which grating was used for each object.
The data reduction has been partially done by using the instrument pipeline\footnote{\url{https://github.com/soar-telescope/goodman_pipeline}} including overscan, trim, slit trim, bias and flat corrections. 
For cosmic rays identification and removal, we used an algorithm as described by \citet{wojtek2004}, which is embedded in the pipeline. 
The extraction and calibration of the spectra were carried out similarly as for Dupont@B\&C using standard \textit{PyRAF} tasks.

\section{Periods and data analysis}
\label{section4}

\begin{figure*}
	\includegraphics[width=\textwidth]{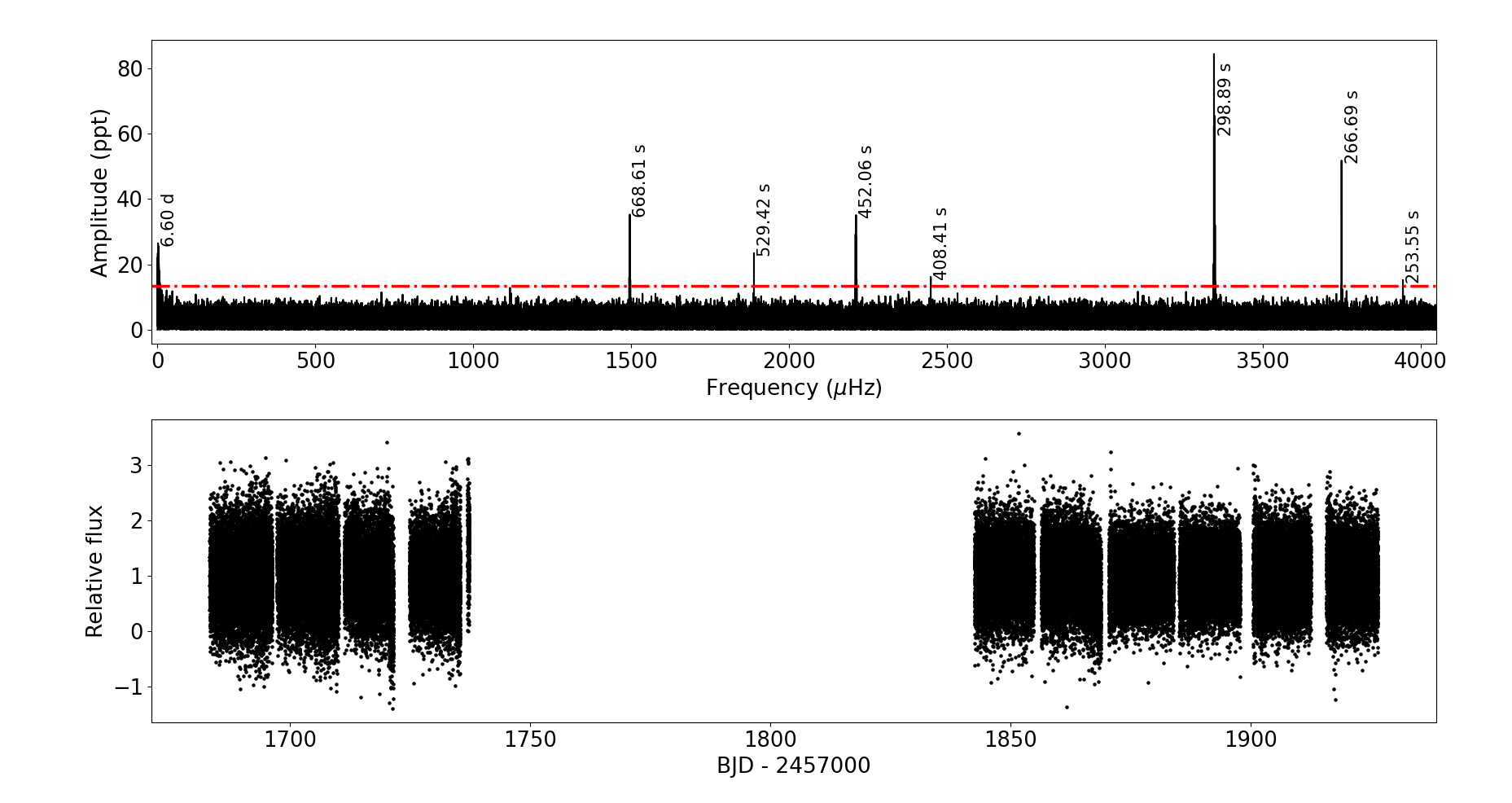}
    \caption{Fourier transform for TIC~313109945, for data with 120~s cadence corresponding to Sectors 14, 15, 20 and 21. The horizontal red line corresponds to the false-alarm probability FAP=1/1000 detection limit for the combined data.}
    \label{FT1}
\end{figure*}

In this section we present the results from the light curve analysis of the 74 new bright ZZ Ceti stars.  The values of the detected periods are listed in Table~\ref{tab:list2} for the results based on TESS data. We also include the corresponding sectors and the amplitude detection limit for false-alarm probability FAP=1/1000. Figure~\ref{FT1} shows the Fourier transform (top panel) and the complete light curve (bottom panel) for the object TIC~313109945. This object shows seven peaks above the FAP=1/1000 confidence level, depicted as a red line. A long period of 6.60~d is also present at low frequencies.

\begin{table*}
    \centering
        \caption{Detected periods for the new ZZ Cetis from TESS. For each object we list the sectors where the target was observed by the TESS satellite, indicating the 20~s cadence runs with "f'' (col~2), the value of the amplitude detection limit for false-alarm probability FAP(1/1000) (col~3), and the list of periods compatible with stellar pulsations in white dwarfs (col~4). We truncate all periods to two decimals places only because the uncertainties in the theoretical models are of the order of 1~s.
        }
    \begin{tabular}{rccc}
    \hline
    TIC	  &   sector		& FAP(1/1000) [ppt] & $\Pi$ [s] (A [ppt])  \\
    \hline
5624184 & f35-f36 & 4.58 & 503.99s (6.28), 445.92 (4.77), 431.31 (4.66)  \\
7675859 & 25,26 & 9.38 & 353.25  (27.39), 356,09 (14.32), 360.32 (9.33), 798.66 (12.08), 743.44 (11.09) \\
8445665 & 24,25 & 8.78 & 812.76 (18.20), 638.22 (15.05), 1018.48 (8.97), 578.05 (8.92), 356.87 (8.84) \\
13566624 & 34 & 7.01 & 421.87 (8.92), 407.63 (7.83) \\
21187072 & 25,26 & 3.63 & 1076.86 (6.79), 1074.27 (4.99), 1070.74 (3.92) \\
24603397 & 5,32 & 1.94 & 262.65 (3.13) \\
29862344 & 03,f30 & 2.67 & 737.57 (3.37), 857.43 (4.03), 898.93 (2.74), 352.09 (2.22) \\
33717565 & 27-29,32,35-36,39 & 3.58 & 364.92 (10.33), 526.98 (4.47) \\
46847635 & 35 & 8.84 & 415.81 (9.66) \\
55650407 & 11-13,f27-f39 & 0.48 & 320.76 (1.65), 262.46 (7.21), 200.08 (4.42), 126.84 (1.84)\\
63281499 & 01,f28 & 2.94 & 320.52 (8.542), 383.70 (2.93) \\
65144290 & 7,33-34 & 4.87 & 278.172 (7.272)\\
72637474 & 03,f30 & 2.51 & 901.16 (2.69), 814.44 (2.56), 966.24 (2.55) \\
79353860 & 1,27 & 3.46 & 945.19 (4.29), 842.43 (3.50), 525.56 (3.60) \\
116373308 & 18 & 40.82 & 361.81 (59.30)  \\
141976247 & 1-8,10-13,27-34,f35-f37 & 0.71 & 261.72 (0.86) \\ 
149863849 & f39 & 1.78 & 397.98 (6.90), 397.04 (6.43), 491.21 (7.45), 568.09 (3.14), 487.36 (3.07) \\
156064657 & 29 & 6.04 & 1418.05 (13.45), 1546.56 (6.05) \\
158068117 & 5-7,32-33 & 1.98 & 268.45 (2.48) \\
167486543 & f32 & 6.27 & 535.26 (19.83), 534.59  (19.55), 535.60 (9.77), 267.29 (7.90) \\
188087204 & 36 & 6.83 & 742.55 (14.44), 657.53 (13.31), 661.35 (9.66), 541.35(7.29), 500.84 (6.87)\\
207206751 & 29-30 & 0.93 & 893.48 (2.07), 776.31 (2.03), 627.26 (1.49), 850.45 (1.47), 906.07 (1.38), \\
   $\cdots$         &  $\cdots$ &  $\cdots$     &1220.01 (1.30), 1270.88 (1.18), 810.13 (1.14), 864.16 (1.13), 939.35 (0.93)\\
220555122 & 1-3,28,30 & 2.23 & 243.89 (2.52), 539.45 (2.36), 137.54 (2.29) \\ 
229581336 & 14-25 & 2.20 & 1106.46 (3.33), 519.02 (2.29), 420.18 (2.20) \\
230029140 & 14-26 & 3.16 & 288.89 (8.63), 311.06 (9.01), 784.77 (7.19), 400.35 (4.64), 364.41 (4.36) \\
230384389 & 14-26 & 0.75 & 457.17 (2.46), 707.92 (2.57), 493.86 (1.67), 749.61 (1.35), 1633.57 (1.05), 1284.58 (0.66) \\
231277791 & 29-f30 & 2.80 & 711.54 (7.03), 497.77 (5.52), 721.02 (5.14), 500.65 (3.64), 717.44 (3.33), 750.86 (3.29), \\
  $\cdots$      &  $\cdots$         &   $\cdots$    & 762.96 (3.20), 719.40 (3.20), 722.74 (2.85), 767.79 (2.80) \\
232979174 & 14-16,21-23 & 2.34 & 282.66 (3.54s) \\
238815671 & 01,f27-28 & 4.37 & 257.59 (9.22), 287.29 (6.86)\\
261400271 & 1,4,7,8,11-13,f27-f28,f31,f34 & 0.70 & 3052.55 (1.46), 295.70 (0.75), 382.92 (0.73)\\
273206673 & 19 & 13.15 & 583.44 (33.65), 827.17 (32.74), 698.89 (19.33), 746.68 (18.62), 892.78 (18.60), 464.40 (16.13),\\ 
    $\cdots$  &  $\cdots$    &   $\cdots$     & 844.06 (15.22), 511.24 (14.72), 688.93 (14.59), 663.24 (14.49), 874.59 (14.24) \\
282783760 & 23 & 6.43 & 257.76 (8.92) \\
287926830 & 15 & 8.34 & 316.22 (9.24) \\
304024058 & 10-11 & 5.30 & 623.43 (2.12), 579.48 (4.39), 506.192 (2.62), 400.28 (2.47)\\
313109945 & 14,15,20-22 & 13.39 & 298.89 (84.01), 266.69 (51.83), 452.06 (34.62), 668.61 (33.33), 529.42 (23.81),408.41 (16.52),\\
 $\cdots$          &    $\cdots$         &   $\cdots$    &253.55 (14.86) \\ 
317153172 & 27,f39 &  5.09  & 786.78 (7.84), 791.96 (6.40, 512.05 (6.53) \\ 
317620456 & 26 & 6.07 & 15601.32 (10.00), 261.10 (7.44), 429.23 (6.16), 2429.96 (6.23) \\
343296348 & 12,13 & 9.92 & 288.27 (16.66), 287.76 (10.01), 287.26 (10.48) \\
344130696 & 12-13,f39 &  2.00 & 1018.68 (2.43), 1057.48 (2.17) \\
353727306 & 18-19,25 & 9.83 & 545.81 (45.43), 470.23 (18.70), 404.84 (16.06), 875.58 (15.32), 463.55 (10.47) \\
370239521 & 1 & 0.52 & 820.43 (1.86), 809.26 (1.18), 575.76 (0.83), 563.65 (0.61), 297.24 (0.53) \\
380298520 & 14,15 & 9.13 & 550.40 (14.68)\\ 
394015496 & 01,27-f28 & 2.26 & 310.27 (4.64), 309.79 (5.38), 309.31 (4.08) \\
415337224 & 5 & 7.56 & 936.55 (15.51), 550.81 (10.97), 953.711 (8.78) \\
428670887 & 10 & 9.19 & 298.14 (17.40) \\
441500792 & 31 & 5.77 & 980.52 (8.21), 786.55 (7.66), 618.20 (@6.52) \\
442962289 & 32 & 6.37 & 481.28 (16.40), 653.63 (13.12) , 498.72 (7.83) \\
610337553 & 30 & 14.19 & 759.60 (32.36), 922.68 (15.42) \\
631161222 & 27,29 & 7.55 & 679.82 (27.33), 708.00 (13.41), 403.63 (11.98), 466.62 (10.40), 367.44 (10.18) \\
631344957 & 28,29 & 6.910 & 363.14 (9.012) \\ 
632543879 & 31 & 13.00 & 461.33 (18.61), 784.54 (16.25), 735.89 (16.08), 652.31 (14.10), 736.24 (13.04) \\
651462582 & 31 & 11.55 & 818.15 (17.55), 683.85 (13.73), 1018.36 (11.88) \\
661119673 & 19 & 42.85 & 626.41 (82.36)\\
683837451 & 27-35,37-39 &  5.69 & 1036.28 (6.11) \\
685410570 & 31-32 & 7.36 & 965.50 (10.23), 812.91 (9.05), 556.66 (7.41)\\
686044219 & 31-32 & 7.87 & 913.00 (18.75), 875.82 (8.95), 736.04 (17.36)\\
\hline
    \end{tabular}
    \label{tab:list2}
\end{table*}

\begin{table*}
\centering
\contcaption{}
	\begin{tabular}{rccc} 
    \hline
    TIC	  &   sector		& FAP(1/1000) [ppt] & $\Pi$ [s] (A [ppt])  \\
    \hline
712406809 & f33 & 7.89 & 828.22 (14.61), 510.29 (11.16), 873.00 (9.61), 115.924 (8.65), 624.28 (7.92) \\ 
724128806 & 27-38,31,34 & 9.15 & 290.18 (10.34) \\
733030384 & 27-29,31-36 & 5.91 & 275.49 (7.35), 411.28 (6.26) \\ 
800153845 & 34 & 16.66 & 878.34 (42.93), 712.44 (17.89) \\
804835539 & 37 & 13.66 & 1007.21 (15.73)\\
804899734 & 30,33,37-39 & 11.18 & 394.71 (24.02), {\it 558.78 (11.31)}\\
951016050 & 37 & 10.39 & 818.45 (15.06), 644.63 (10.63) \\
1001545355 & 14,20-22 & 10.60 & 516.18 (17.79), 761.14 (15.78), 955.19 (15.47), 1058.10 (11.08), 249.97 (10.99) \\
1102242692 & 16,22,24 & 7.15 & 1009.04 (9.56), 406.17 (7.37) \\
1102346472 & 15-16,22-23 & 11.08 & 458.13 (27.73) \\
1173423962 & 38 & 28.32 & 618.36 (38.36), 794.80 (30.33) \\
1108505075 & 39 & 34.74 & 693.55 (45.47), 1323.54 (36.69), 1801.77 (36.09) \\ 
1201194272 & 14-16,18-26 & 6.90 & 840.92 (8.37) \\
1309155088 & 25 & 12.94 & 769.06 (17.29) \\
1989258883 & 27 & 7.29 & 909.04 (10.78)  \\
1989866634 & 27 & 14.28 & 613.97 (22.22) \\
2026445610 & 27 & 14.69 & 825.25 (25.56), 815.53 (20.52), 317.74 (15.17) \\
2055504010 & 28 & 12.81 & 990.32 (14.56), 818.73 (12.92), 774.69 (13.25) \\
\hline
 \end{tabular}
\end{table*}

The results from ground based observations are presented in Table~\ref{tab:GB-obs}. From the 11 objects observed at higher cadence from ground-based telescopes, eight of them show periodic variability above the FAP=1/1000 limit on the ground data. For TIC~55650407 we confirm some periods detected from TESS data, but no additional periods were detected, as can be seen from Figure~\ref{TESS-SOAR}. This is also the case for TIC~2055504010. For TIC~304024058 the same periods as in the TESS data were detected from ground-base observations, as shown in Figure~\ref{Nyquist}.

For TIC~273206673, TIC~282783760, and TIC~1989866634 we detected additional periodicities from ground-based observations, and also confirmed the periods identified by using TESS data. In particular, for TIC~370239521, the period spectrum present in the FT from the Pico dos Dias observatory are not the same periods detected by TESS data; however, the periods are in the same range around $\sim 800$ s. Only one period is present in both data sets, with a period of $\approx 279$ s. 

\begin{table}
	\centering
	\caption{Detected periods from ground based observations. The TIC number, period, amplitude are listed in columns 1, 2 and 3, respectively. The identification for each mode is listed in the last column.}
	\label{tab:GB-obs}
	\begin{tabular}{rccc} 
\hline
TIC & $\Pi$ [s] & A (ppt) & ID\\
\hline 
20979953 & 259.68 & 7.36 & $f_1$ \\ 
           & 285.30 & 6.25 & $f_2$\\
           & 365.64 & 4.07 &  $f_3$\\ \hline
55650407 & 262.82 &  12.78 & $f_1$ \\
           & 199.72 &  7.78 & $f_2$\\ \hline
273206673 & 1029.00  & 49.00 & $f_1$\\\hline
282783760 & 257.59 & 2.20 & $f_1$\\
           & 283.42 & 1.30 & $f_2$\\
           & 308.96 & 1.5 & $f_3$\\ \hline
304024058 & 620.03 & 13.53 & $f_1$\\
           & 577.37 & 16.75 & $f_2$\\
           & 505.36 & 18.42 & $f_3$\\
           & 399.64 & 10.60 & $f_4$\\ \hline
370239521 & 894.90 & 18.68 & $f_1$ \\
           & 733.53 & 10.65 & $f_2$\\
           & 447.21 & 7.05 & 2$f_1$ \\
           & 776.76 & 6.93 & $f_3$\\
           & 934.29 & 6.51 & $f_4$ \\
           & 297.61 & 6.32 & 3$f_1$ \\  \hline
1989866634 & 613.96 & 55.86 & $f_1$ \\
           & 305.92 & 15.10 & $2f_1$\\
           & 570.39 & 14.70 & $f_2$\\
           & 362.83 & 14.21 & $f_3$\\
           & 896.05 & 12.02 & $f_4$\\
           & 228.65 & 10.54 & $f_5$\\
           & 501.38 & 7.96 & $f_6$\\
           & 294.37 & 7.38 & $f_1 + f_2$ \\
           & 975.11 & 7.26 & $f_7$\\ \hline
2055504010 & 772.94 & & $f_1$ \\
\hline 
\end{tabular}
\end{table}

\subsection{Super-Nyquist}

TESS acquires data with an even time sampling between observations, $\Delta t$. Even time sampling causes each significant signal to produce an infinite set of alias signals in the periodogram, each reflected across multiples of the Nyquist frequency of $f_{\rm Nyq} = 1/(2\Delta t)$.  Without external constraints on intrinsic signal frequencies, any of these frequency aliases can describe the data equally well. Deviations from strictly even time series caused by the barycentric timestamp corrections are minor \citep{2015MNRAS.453.2569M}, and strong aliasing is observed in the TESS data.

With pulsation periods observed to be as short as 70\,s, ZZ Cetis can pulsate with intrinsic signals above up to three times the 120~s cadence Nyquist frequency of 4166.67\,$\mu$Hz, resulting in up to four viable aliases for each pulsation mode. Including an incorrect alias frequency in an asteroseismic analysis will corrupt the inference. Without additional data, frequency aliases are usually favoured that appear most consistent with the ensemble of studied ZZ Ceti pulsation spectra. In some cases, coherence of a pulsation signal can be used as an argument for certain aliases, and modes with periods longer than roughly 800~s tend to vary in phase and amplitude on $\sim$day timescales \citep{2017ApJS..232...23H,2020ApJ...890...11M}.

From Sector 27 on, the TESS satellite observed some objects with 20~s cadences, effecting a Nyquist frequency sufficiently above ZZ Ceti pulsation frequencies to avoid Nyquist ambiguities. For targets without such data, ground-based follow-up photometry with a different cadence can be used to select the correct alias \citep[e.g.,][]{2017ApJ...851...24B}. Based on the number of targets with an f in Table~\ref{tab:list2} representing the faster observing cadence, {\bf 15} of our {\bf 74} targets have 20-s cadence from TESS.  

\begin{figure*}
	\includegraphics[width=\textwidth]{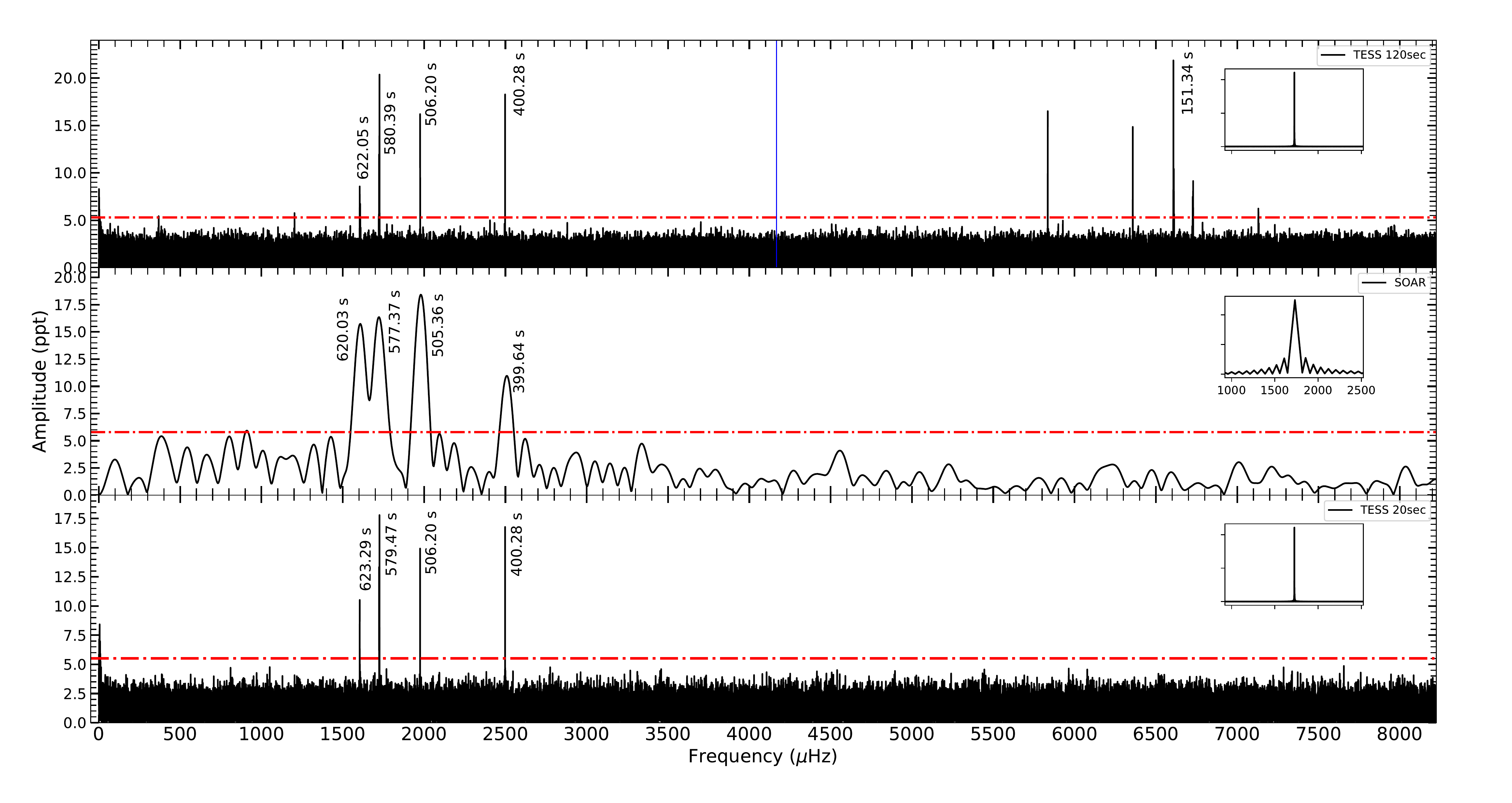}
    \caption{Fourier transform for TIC~304024058 for three different data sets. The result based on the 2-min (120-s) cadence TESS data is depicted in the top panel, while the bottom panel shows the FT for the 20-s cadence TESS data. The FT based on the SOAR telescope data (10 s integration time) is shown in the middle panel. The amplitude at FAP=1/1000 detection limit in each case is indicated by the horizontal dashed line, and the spectral window for each case is depicted as an inset plot. The blue vertical line in the top panel indicates the value of the Nyquist frequency.}
    \label{Nyquist}
\end{figure*}

\begin{figure*}
	\includegraphics[width=\textwidth]{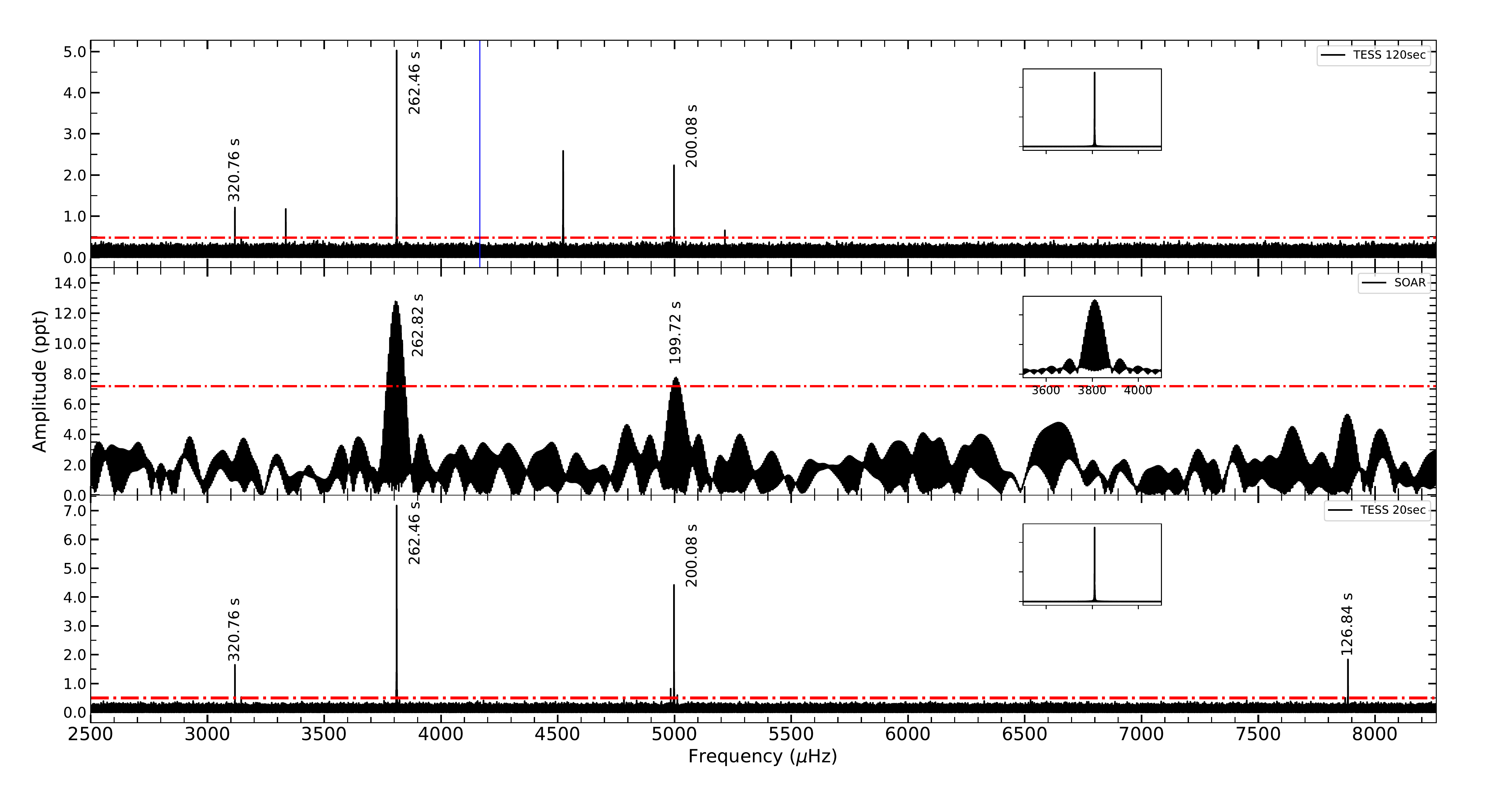}
    \caption{Fourier transforms for TIC~55650407 (same as Fig. \ref{Nyquist}).}

    \label{TESS-SOAR}
\end{figure*}

As an example, in Figure~\ref{Nyquist} we show the FT for TIC~304024058 for 120~s cadence TESS data (top panel), ground-based observation from SOAR telescope with $\sim$15~s-cadence (middle panel) and 20~s cadence TESS data (bootm panel). The red horizontal line indicates the detection limit and the inset in each plot depicts the spectral window. For the 120~s cadence data in the top panel, the blue vertical line corresponds to the Nyquist frequency. Note that in this case, there are four significant peaks in the super-Nyquist region, with the peak at a period of 151.34~s being the one with the highest amplitude. However, these peaks are not seen in the ground-based observations, where only the peaks corresponding to periods between 400~s and 620~s are present. The same result is obtained when we include the 20~s cadence data from TESS. Thus, in this case, the sub-Nyquist aliases are confirmed to correspond to the intrinsic pulsation frequencies by ground-based observations and shorter-cadence TESS data. 

A different result is found for TIC~55650407, as shown in Figure~\ref{TESS-SOAR}. The FT corresponding to the 120~s cadence TESS data shows two peaks with super-Nyquist frequencies, in particular one corresponding to a period of 200.08~s (see top panel). The same period is also present in the FT for the SOAR telescope data, shown in the middle panel of Figure~\ref{TESS-SOAR}, which confirms the existence of this period. On the other hand, the peak corresponding to the period of 320.76~s is not detected in the SOAR data, probably due to the much shorter SOAR observation run (3.72 h). Finally, the period with 200.08~s is also confirmed by the 20~s cadence TESS data, shown in the bottom panel of Figure \ref{TESS-SOAR}. In addition, there is another peak at high frequencies with a corresponding period of 126.84~s that was not apparent in the 120-s data since the period is close to the exposure time.

\subsection{TIC~20979953}
\label{new-tic}

\begin{figure}
	\includegraphics[width=0.50\textwidth]{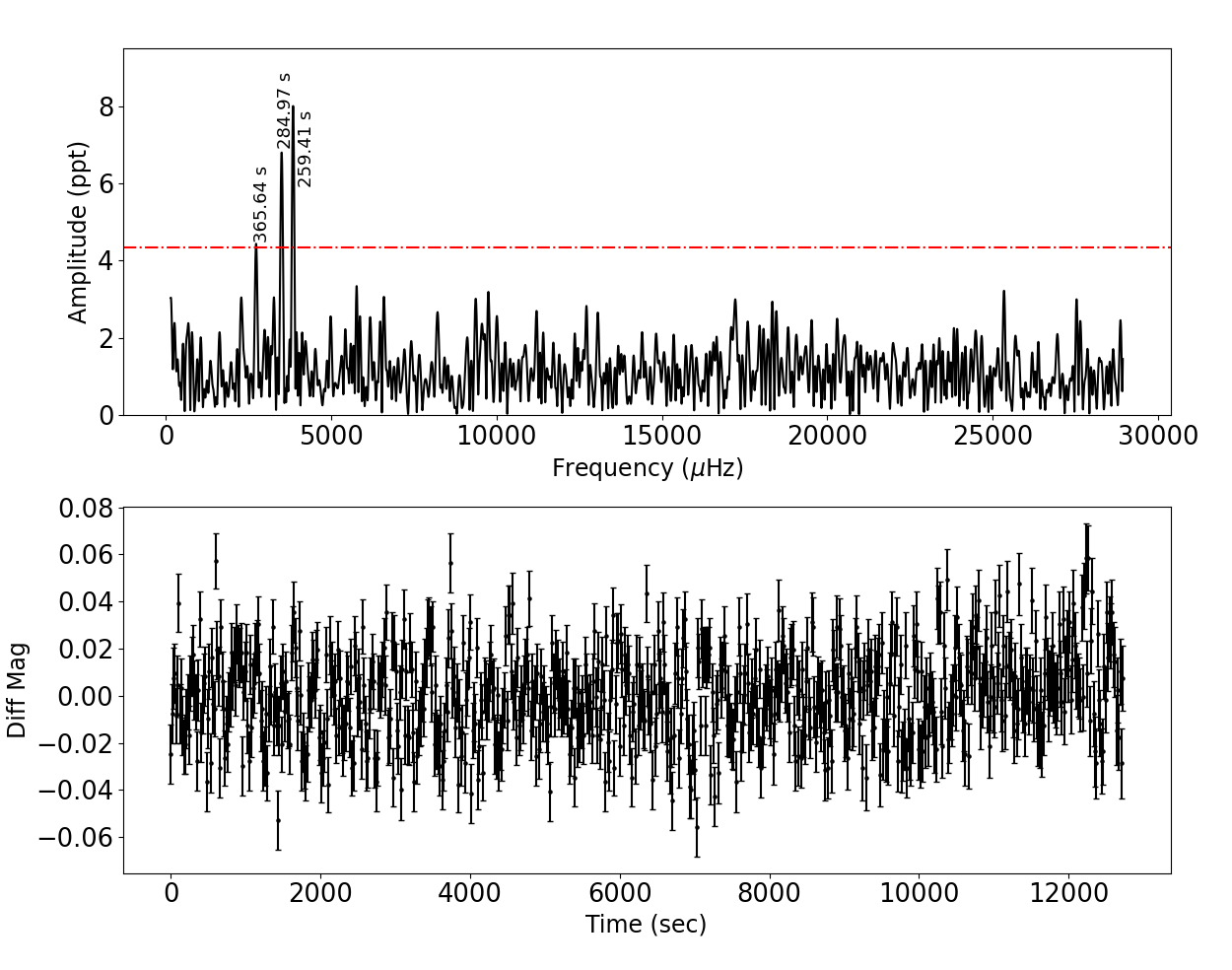}
    \caption{FT and light curve, top and bottom panel respectively, for TIC~20976653 based on data obtained from the Pico dos Dias observatory on 2020-06-14. The data spans for 3.4~h, with an integration time of 17~s.}
    \label{TIC0020}
\end{figure}

TIC~20979953 was part of the target list for the TESS mission, but no 120~s or 20~s cadence data was taken for this object to date. We observed TIC~20979953 from the Pico do Dias observatory in two nights in 2020 (see Table~\ref{tab:GB}) for a total of 6.76~h. The light curve and FT are presented in Figure \ref{TIC0020} for the night of 2020-06-14 that spans for 3.4~h. From this dataset we detected three short periods of 259.68~s, 285.30~s and 365.64~s, compatible with a blue edge ZZ Ceti \citep{2006ApJ...640..956M}. 

\section{Asteroseismological analysis}
\label{section5}

In this section we present an asteroseismological analysis of all objects presented in Tables~\ref{tab:list2} and \ref{tab:GB-obs}. 
We employed an updated grid of DA white dwarf models, obtained from full evolutionary computations of the progenitor star. They were generated using the LPCODE evolutionary code \citep[see][for details]{2010ApJ...717..897A,2010ApJ...717..183R,2015MNRAS.450.3708R}, that computes the evolution of the star from the zero-age main sequence, considering hydrogen and helium central burning and the giant phases. The grid corresponds to C/O-core white dwarf models with stellar masses between 0.493~M$_{\odot}$ to 1.05~M$_{\odot}$, with a hydrogen layer mass in the range of $\sim 4\times 10^{-4}$ to $\sim 10^{-10}$~M$_*$, depending on the stellar mass. This model grid is an extended version of the model grid presented in \citet{2019MNRAS.490.1803R} that includes new cooling sequences with stellar masses between 0.5 and 1.0~M$_{\odot}$, 
along with approximately eight hydrogen layer values for each sequence, depending on the stellar mass. For hydrogen envelopes thinner than $10^{-10}$M$_*$ the outer convective zone will mix the hydrogen into the more massive helium layer before reaching the ZZ Ceti instabily strip, turning the star into a DB white dwarf \citep{2020MNRAS.492.5003O, 2020MNRAS.492.3540C}.
For each model in our grid we computed non-radial g-mode pulsations, considering the adiabatic approximation, using the adiabatic version of the LP-PUL pulsation code \citep[see][for details]{2006A&A...454..863C}.

For each object we search for an asteroseismological representative model that best matches the observed periods. To this end, we seek for the theoretical model that minimizes the quality function given by \citep{2008MNRAS.385..430C},

\begin{equation}
    S(M_*, M_{\rm H}, T_{\rm eff})=\sqrt{\sum_{i=1}^N{\rm min}\left[\frac{[\Pi_i^{\rm th}-\Pi_i^{\rm obs}]^2 A_i}{\sum^{N}_{i=1} A_i}\right]},
\end{equation}

\noindent where $N$ is the number of observed periods, $\Pi^{\rm th}_i$ is the theoretical period that better fits the observed period $\Pi^{\rm obs}_i$, and the amplitudes $A_i$ are used as weights for each period. In this way, the period fit is more influenced by those modes with larger observed amplitudes. We also compute other quality functions without the amplitude weighting, obtaining similar results.

For the eight objects with both TESS and ground-based data, we combine (add) the list of periods to perform the asteroseismological fit. We do not consider periods corresponding to harmonics or linear combinations, nor those modes that show super-Nyquist frequencies that were not confirmed by short-cadence observations. In the case where a period is detected both from TESS and ground-based observations, we consider the period value and amplitude corresponding to the TESS data. 
For the stars with spectroscopic mass below the minimum value of our C/O-core grid (0.493~M$_{\odot}$) we also perform an initial asteroseismological fit with He-core white dwarf models with stellar masses from 0.17 to 0.45~M$_{\odot}$ \citep{2012A&A...547A..96C}, considering only canonical hydrogen envelopes and $\ell =1$ modes. Finally, if necessary, we consider the values of effective temperature and mass from Table \ref{tab:list1} as an additional constrain in the fitting procedure. These last constrains are particularly important for objects with only one or two detected periods, which is the case for a substantial fraction of our objects. However, note that some particular short ($\lesssim$ 200 sec) periods can be strong constraints on their own, since they propagate in the inner parts of the star and are particularly sensitive to the inner structure.

The results of our asteroseimological fits are presented in Table~\ref{tab:list1-res}. For each object we list the stellar mass, thickness of the hydrogen envelope and effective temperature for the seismological model, in columns 2, 3, and 4, respectively. Column 5 shows the values for the theoretical periods along with the corresponding harmonic degree $\ell$, and radial order $k$. Finally, the value of the quality function $S$ is listed in column 9. 
The first model listed is the one we choose to be the best-fitting model for that particular object.

\begin{table*}
	\centering
	\caption{Best fit model for the new ZZ Cetis using the list of observed modes. The stellar mass, hydrogen envelope and effective temperature are listed in columns 2, 3 and 4, respectively. We list the theoretical periods in column 5, along with the harmonic degree and the radial order. The value of the quality function $S$ in seconds is listed in column 6. }
	\label{tab:list1-res}
	\begin{tabular}{rccccc} 
\hline
TIC   &  M/M$_{\odot}$ & $-\log($M$_{\rm H}$/M$_*)$ & T$_{\rm eff}$ [K] &  $\Pi$[s] ($\ell, k$) & $S$ [s] \\
\hline
5624184 & 0.493 & 4.85 & 11190 & 503.71 (1,7), 446.70 (2,12), 431.35 (1,16) & 0.46 \\
7675859 & 0.660 & 8.82 & 11710 & 356.26 (1,4), 798.51 (1,12), 742.93 (1,13) & 0.28 \\
  $\cdots$ & 0.800 & 5.67 & 11660 & 356.33 (1,6), 799.19 (1,18), 743.00 (2,30) & 0.37 \\  
8445665 & 0.675 & 4.35 & 10970 & 811.70 (1,17), 639.25 (1,13), 358.27 (2,12), 575.46 (2,21), 1019.48 (2,39) & 1.51 \\
13566624 & 0.705 & 6.15 & 12770 & 421.87 (1,4) & $\cdots$ \\ 
21187072 & 0.660 & 5.55 & 11680 & 1074.2711 (1,21) & $\cdots$ \\ 
24603397 & 0.542 & 6.83 & 12500 & 262.65 (1,2) & $\cdots$ \\ 
29862344 & 0.609 & 8.33 & 11370 & 737.29 (1,12), 857.39 (1,14), 89890 (2,27) & 0.12 \\
33717565 & 0.609 & 5.74 & 12530 & 262.72 (1,3), 198.70 (1,2), 322.30 (2,9) & 0.99 \\
46847635 & 0.686 & 4.87 & 11930 & 415.82 (1,7) &  $\cdots$ \\ 
63281499 & 0.542 & 4.25 & 11610 & 320.37 (1,4), 384.76 (1,6) & 0.60 \\ 
65144290 & 0.632 & 4.46 & 11480 & 278.17 (1,4) & $\cdots$ \\  
72637474 & 0.542 & 4.94 & 11720 & 812.13 (1,14), 901.66 (1,16), 966.44 (1,17) & 1.36 \\
79353860 & 0.686 & 5.25 & 11390 & 945.54 (1,18), 842.57 (1,16), 525.10 (1,9) & 0.35\\
116373308 & 0.609 & 8.33 & 11590 & 361.81 (1,4) & $\cdots$ \\
$\cdots$   & 0.609 & 5.54 & 12160 & 361.78 (1,5) & $\cdots$ \\ 
141976247 & 0.686 & 8.82 & 12910 & 261.71 (1,3) & $\cdots$ \\
149863849 & 0.660 & 4.41 & 11380 & 397.99 (2,13), 419.15 (2,14), 569.41 (2,20), 487.83 (2,17) & 0.55 \\
156064657 & 0.493 & 6.84 & 10860 & 1418.07 (1,22), 1546.64 (1,24) & 0.05 \\
$\cdots$   & 0.358 & 3.26 &  9640 & 1418.71 (1,20), 1547.54 (1,22) & 1.04 \\
158068117 & 0.493 & 8.82 & 12150 & 268.453 (1,2) & $\cdots$ \\ 
$\cdots$   & 0.303 & 2.90 & 9350  & 268.436 (1,2) & $\cdots$ \\
167486543 & 0.820 & 4.93 & 12700 & 267.27 (1,5), 535.10 (1,13) & 0.13  \\
$\cdots$        & 0.745 & 5.37 & 12230 & 267.06 (1,4), 535.29 (1,11) & 0.17\\
188087204 & 0.493 & 4.16 & 10640 & 743.05 (1,12), 657.69 (1,8), 544.00 (1,8), 500.07 (2,14) & 1.19 \\ 
207206751 & 0.570 & 4.28 & 10950 & 894.12 (2,30), 775.00 (2,26), 859.69 (1,16), 626.64 (1,11), 905.59 (1,17) & 2.55 \\
           &       &      &       & 1115.36 (2,38), 1277.72 (1,25), 864.29 (2,29), 809.88 (1,15) &  \\
220555122 & 0.686 & 6.34 & 11690 & 243.908 (1,3), 539.408 (1,9) & 0.03 \\
229581336 & 0.493 & 4.45 & 11310 & 1106.45 (2,34), 519.59 (2,15), 420.19 (1,6) & 0.31 \\
 $\cdots$  & 0.400 & 3.18 & 10460 & 1106.60 (1,17), 514.61 (1,7), 421.31 (1,5) & 1.75 \\          
230029140 & 0.593 & 5.04 & 11190 & 287.01 (1,3), 313.76 (1,4), 784.77 (1,14), 400.97 (1,6), 360.26 (2,10) & 2.26 \\
230384389 & 0.525 & 9.25 & 11410 & (457.61 (1,5), (708.73 (1,10), 495.23 (2,12), 751.67 (2,20), & 1.24\\
           &       &      &       &  1283.10 (1,20), 1632.01 (2,46) & \\
231277791 & 0.570 & 5.45 & 11300 & 721.23 (1,13), 713.02 (2,23), 498.55 (1,8), 762.44 (2,25) & 0.86  \\
232979174 & 0.660 & 5.35 & 12020 & 282.66 (1,4) & $\cdots$ \\
$\cdots$   & 0.493 & 3.72 & 11710 & 282.67 (1,3) & $\cdots$ \\
238815671 & 0.690 & 5.26 & 11630 & 257.792 (1,3), 286.891 (1,4) & 0.30\\ 
261400271 & 0.820 & 5.78 & 12390 & 295.25 (1,5), 382.68 (1,7) & 0.36\\
287926830 & 0.570 & 4.28 & 11320 & 316.23 (1,4) & $\cdots$ \\
313109945 & 0.675 & 9.25 & 9890 & 300.19 (1,3), 266.11 (1,2), 450.54 (2,11), 685.84 (2,19), 583.53 (2,16),  & 1.74 \\
           &       &      &       & 410.63 (2,10), 250.03 (2,5)  & \\  
317153172 & 0.621 & 6.34 & 11900 & 786.67 (2,25), 791.88 ( 1,14), 512.27 (1,8) & 0.15 \\ 
317620456 & 0.632 & 4.46 & 11010 & 260.86 (1,3), 429.24 (1,7) & 0.19\\
343296348 & 0.548 & 4.27 & 11310 & 287.766 (1,3) & $\cdots$ \\ 
344130696 & 0.632 & 9.34 & 11180 & 1018.70 (1,17), 1057.84 (1,18) & 0.31 \\
345202693 & 0.705 & 4.88 & 10670 & 587.86 (1,11), 789.31 (1,16), 833.79 (1,17), 833.79 (1,17) & 0.51 \\
353729306 & 0.690 & 6.94 & 11680 & 545.96 (1,9), 470.63 (1,7), 404.13 (2,12), 875.57 (2,30) & 0.38 \\  
380298520 & 0.745 & 9.24 & 11550 & 549.86 (1,9) & $\cdots$ \\ 
394015496 & 0.593 & 6.11 & 11570 & 309.79 (1,3) & $\cdots$ \\ 
415337224 & 0.609 & 4.85 & 10100 & 936.88 (2, 32), 550.73 (1,9), 953.11 (1,18) & 0.37 \\
428670887 & 0.609 & 5.24 & 11500 & 298.14 (1,4) & $\cdots$ \\  
441500792 & 0.705 & 4.48 & 11260 & 617.31 (1,13), 786.32 (1,17), 980.86 (1,22) & 0.53 \\
$\cdots$ & 0.745 & 9.28 & 11460 & 618.21 (2,20), 786.52 (2,26), 981.21 (2,33) & 0.41 \\
442962289 & 0.837 & 5.00 & 12120  & 418.75 (1,9), 652.00 (1,16), 498.87 (2,22) & 1.02\\
610337553 & 0.609 & 6.33 & 10970 & 759.49 (1,13), 922.81 (1,16) & 0.12 \\
631161222 & 0.609 & 5.44 & 11400 & 368.60 (1,5), 403.39 (1,6), 467.10 (1,7), .79,27 (1,12), 708.31 (1,13) & 0.60 \\
631344957 & 0.550 & 4.84 & 11550 & 363.144 (1,5) & $\cdots$ \\
632543979 & 0.660 & 5.15 & 11250 & 461.54 (2,15), 783.90 (1,15), 736.26 (2,25), 652.00 (2,22), 797,03 (1,14) & 0.50 \\
651462582 & 0.593 & 5.79 & 10780 & 817.49 (1,14), 683.29 (1,11), 1019.36 (1,18) & 0.74 \\
661119673 & 0.570 & 4.55 & 11600 & 626.42 (1,11) & $\cdots$ \\ 
685410570 & 0.609 & 4.95 & 10900 & 965.62 (1,18), 812.60 (1,15), 557.09 (1,9) & 0.30 \\ 
686044219 & 0.639 & 4.12 & 11130 & 913.75 (1,19), 735.99 (1,15), 875.08 (1,18) & 0.58 \\
\hline
\end{tabular}
\end{table*}

\begin{table*}
\centering
\contcaption{}
	\begin{tabular}{rccccc} 
\hline
TIC   &  M/M$_{\odot}$ &  M$_{\rm H}$/M$_*$ & T$_{\rm eff}$ [K] &  $\Pi$[s] ($\ell, k$) & $S$ [s] \\
\hline
712406809 & 0.646 & 4.12 & 10820 & 827.83 (1,17), 510.53 (2,18), 872.93 (1,18), 118.57 (1,1), 623.51 (1,12) & 1.15 \\ 
724128806 & 0.542 & 6.36 & 10910 & 290.18 (1,3) & $\cdots$ \\ 
$\cdots$   & 0.251 & 2.92 & 10410 & 290.19 (1,2) & $\cdots$ \\
733030384 & 0.660 & 6.24 & 12390 & 275.48 (1,4), 411.63 (1,6) & 0.24 \\
800153845 & 0.593 & 7.34 & 11780 & 877.99 (1,14), 712.38 (1,11) & 0.29 \\ 
804835539 & 0.609 & 4.45 & 10990 & 1007.21 (1,20) & $\cdots$  \\
804899734 & 0.609 & 5.35 & 11780 & 394.82 (1,6) & $\cdots$  \\ 
951016050 & 0.660 & 4.86 & 10850 & 818.58 (1,12), 644.56 (1,16) & 0.11 \\
1001545355 & 0.542 & 6.13 & 11380 & 516.47 (1,14), 763.42 (1,22), 955.84 (1,28), 1055.98 (1,31) & 1.52 \\
1102242692 & 0.609 & 5.44 & 11200 & 1009.13 (1,19), 406.18 (1,6) & 0.07 \\
1102346472 & 0.548 & 4.27 & 10970 & 458.13 (1,7) & $\cdots$ \\ 
1108505075 & 0.579 & 5.34 & 11310 & 693.40 (1,12), 1323.63 (1,25), 1801.95 (1,35) & 0.14 \\
1173423962 & 0.690 & 7.14 & 10940 & 618.45 (1,10), 794.75 (1,14) & 0.08 \\
1201194272 & 0.609 & 4.11 & 11470 & 840.92 (1,17) & $\cdots$ \\ 
1309155088 & 0.609 & 4.19 & 10710 & 769.05 (1,15) & $\cdots$ \\ 
1989258883 & 0.609 & 6.04 & 11110 & 909.03 (1,16) & $\cdots$ \\ 
2026445610 & 0.525 & 3.79 & 11240 & 825.42 (1,15), 317.77 (1,4) & 0.09 \\
\hline
\end{tabular}
\end{table*}

\begin{table*}
	\centering
	\caption{Best fit model for the new ZZ Cetis using the list of modes from TESS and/or ground based observations. The stellar mass, hydrogen envelope and effective temperature are listed in columns 2, 3 and 4, respectively. We list the theoretical periods in column 5, along with the harmonic degree and the radial order. The value of the quality function $S$ in seconds is listed in column 6.  *No variability detected with TESS down to FAP=1/1000.}
	\label{tab:listcomb}
	\begin{tabular}{rccccc} 
\hline
TIC   &  M/M$_{\odot}$ & $-\log($M$_{\rm H}$/M$_*)$ & T$_{\rm eff}$ [K] &  $\Pi$[s] ($\ell, k$) & $S$ [s] \\
\hline
20979953 & 0.632 & 8.33 & 11130 & 259.77 (1,2), 285.27 (1,3), 365.57 (1,4) & 0.07\\
           & 0.593 & 3.93 & 12200 & 258.59 (1,3), 285.59 (1,4), 365.79 (1,6) & 0.73 \\ 
55650407 & 0.570 & 3.82 & 12600 & 316.93 (1,5), 262.34 (1,3), 203.39 (2,5), 125.17 (1,1) & 2.28 \\ 
           & 0.542 & 5.63 & 12980 & 320.95 (1,4), 263.45 (1,3), 201.75 (2,4), 125.17 (2,2) & 1.25 \\ 
273206673 & 0.686 & 4.60 & 11170 & 582.89 (1,11), 825.86 (1,17), 693.35 (1,14), 744.84 (2,27), & 2.32 \\
           &       &      &       & 894.33 (2,33), 464.05 (1,8), 841.82 (2,31), 508.83 (2,18),  & \\
           &       &      &       & 688.38 (2,25),665,46 (2,24), 873.64 (1,18), 1033.89 (2,38) &  \\        
282783760 & 0.593 & 3.93 & 12270 & 258.14 (1,3), 284.80 (1,4), 307.69 (1,4) & 1.06 \\
           & 0.493 & 4.35 & 11710 & 257.84 (2,6), 283.94 (1,3), 308.38 (2,8) & 0.45 \\ 
304024058 & 0.542 & 4.15 & 12020 & 623.34 (2,20), 579.09 (1,10), 504.59 (2,16), 400.22 (2,12) & 0.80*\\
           & 0.593 & 6.33 & 11400 & 623.53 (2,19), 578.50 (1,9), 506.82 (2,15), 401.33 (2,11) & 0.83* \\ 
370239521 & 0.770 & 8.66 & 11010 & 822.34 (1,15), 806.91 (2,27), 576.79 (1,10), 565.95 (2,18), 274.34 (2,7), & 2.05 \\
           &       &      &       & 895.47 (1,17), 732.69 (2,24), 779.20 (2,26), 932.76 (1,18) & \\
           & 0.721 & 5.08 & 11240 & 821.03 (2,31), 808.78 (1,17), 578.19 (2,21), 569.39 (1,11), 276.93 (1,4),  & 2.57 \\
           &       &      &       & 896.39 (1,19), 729.89 (1,15), 772.40 (2,29), 932.48 (1,20) & \\      
1989866634 & 0.820 & 7.36 & 10960 & 613.81 (1,12), 568.43 (1,11), 364.26 (2,12), 899.66 (2,33),  & 1.47 \\
           &       &      &       & 227.90 (1,3), 500.76 (1,9), 973.80 (2,36) & \\
           & 0.609 & 4.75 & 11250 & 614.31 (1,11), 572.38 (2,19), 359.98 (2,11), 895.59 (2,31), & 2.03 \\
           &       &      &       &  227.89 (2,6), 495.11 (1,8), 975.76 (1,19)  &   \\ 
2055504010 & 0.705 & 5.75 & 11030 & 990.45 (1,20), 818.45 (1,16), 774.90 (2,27) & 0.21 \\ 
\hline
\end{tabular}
\end{table*}

\subsection{Rotation Periods}

White dwarf stars are considered slow rotators, with rotation periods between a few hours and several days \citep[see for instance][]{2017EPJWC.15201011K}.
By considering the frequency separation we can estimate the rotation period of the white dwarf star, following the equation \citep{1949ApJ...109..149C, 1951ApJ...114..373L}:

\begin{equation}
    \frac{1}{P_{\rm rot}} = \frac{\Delta \nu_{k,\ell, m}}{m (1-C_{k\ell})}
\end{equation}

\noindent where $m$ is the azimuthal number and $C_{k\ell}$ is the rotational splitting coefficient given by:

\begin{equation}
    C_{k,\ell}=\frac{\int_0^{R_*} \rho r^2 [2\xi_r\xi_t + \xi_t^2 dr}{\int_0^{R_*} \rho r^2[\xi_r^2 + \ell(\ell + 1)\xi_t^2]dr}
\end{equation}

\noindent where $\rho$ is the density, $r$ is the radius and $\xi_r$ and $\xi_t$ are the radial and horizontal displacement of the material. 

Since all our datasets have very high duty cycles, our TESS data is generally free of aliasing and can readily reveal patterns of even frequency spacing in the Fourier transform that can reveal the stellar rotation rate \citep[e.g.][]{2004IAUS..215..561K}). Identifying rotationally split multiplets is also an excellent way to identify the spherical degree and azimuthal order of the modes present (e.g., \citealt{1991ApJ...378..326W,1994ApJ...430..839W}).

\begin{figure}
	\centering
	\includegraphics[width=0.9\columnwidth]{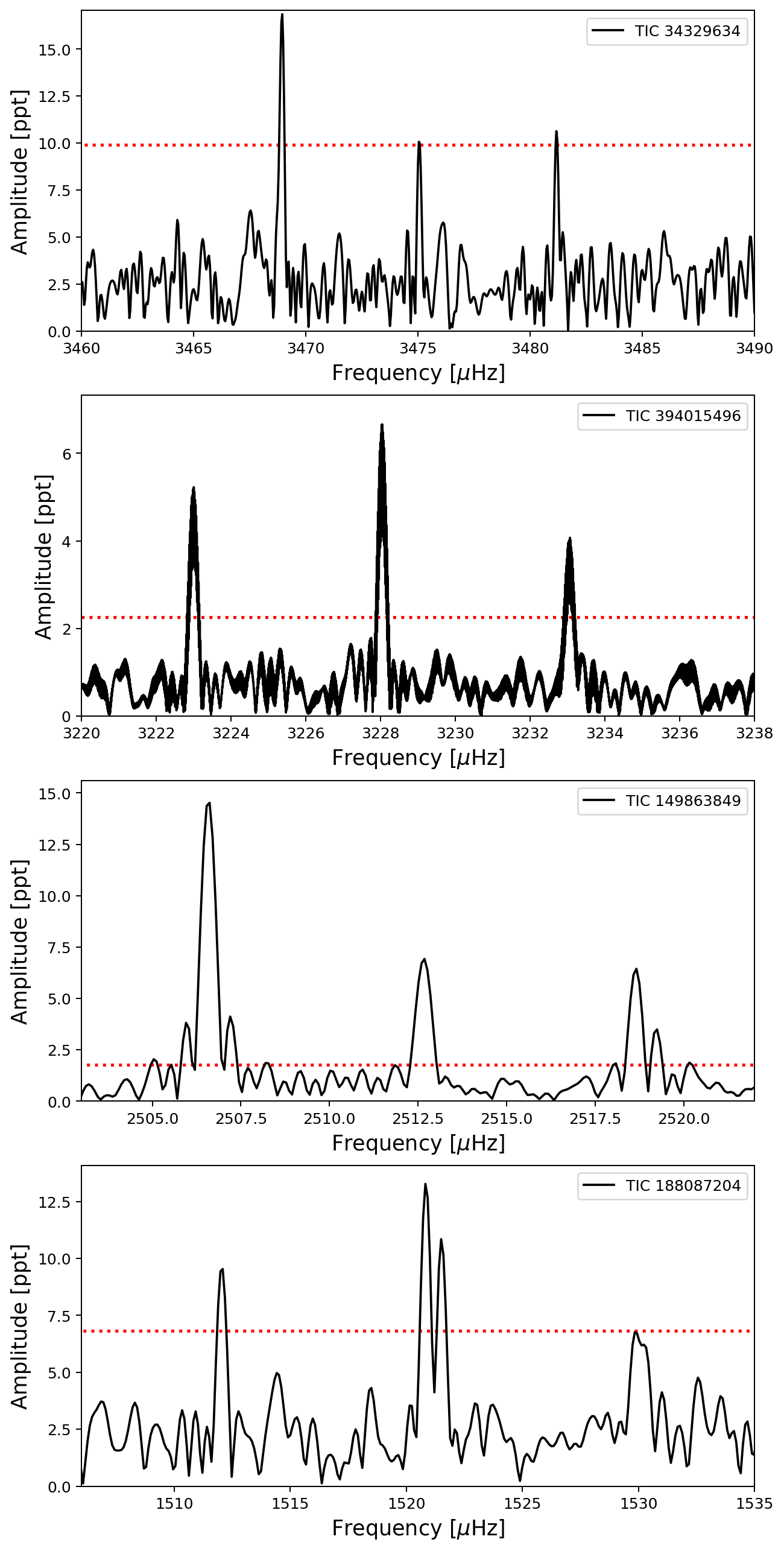}
    \caption{FT showing the three peaks corresponding to the triplet for four objects listed in Table \ref{rota}. }
    \label{triplet}
\end{figure}

We detect rotationally split multiplets in four new ZZ Cetis with TESS data. The list of objects with detected $\ell=1$ multiplets is presented in Table~\ref{rota}. 
We list the frequencies corresponding to the multiplet, the $C_{k\ell}$ (determined from our best-fit asteroseismic solution in Section~\ref{section5}), and the resultant mean stellar rotation period. In Figure~\ref{triplet} we show all four stars with rotationally split multiplets; these are the only multiplets we identify in each star.

\begin{table}
	\centering
	\caption{List of objects with detected rotational splittings, also shown in Figure~\ref{triplet}. For each object we list the frequency that form the multiplet (col 2), the value of the $C_{k\ell}$ obtained from the asteroseismologial representative model (col 3) and the mean rotation period (col 4).}
	\label{rota}
	\begin{tabular}{rccc} 
\hline
TIC   &  $\nu$ [$\mu$Hz] & $C_{k\ell}$  &  $\bar{P}_{\rm rot}$ \\
\hline
7675859 & 2830.86, 2808.28, 2775.31 & 0.487 & 5.20 h \\
21187072 & 933.93, 930.86, 928.63 & 0.495 & 2.24 d \\
343296348 & 3481.29, 3475.12, 3468.97 & 0.458 & 1.02 d \\ 
394015496 & 3233.00, 3227.99, 3223.00  & 0.463 & 1.24 d \\
\hline
\end{tabular}
\end{table}

The lack of aliasing from space-based data from Kepler and TESS has enabled the detection of rotational splittings for a much larger sample of pulsating white dwarfs. Among the first 27 ZZ Cetis observed by Kepler and K2, patterns from rotational splitting was observed in 20 ZZ Cetis (74\,per\,cent, \citealt{2017ApJS..232...23H}). Unfortunately, we only detect patterns from rotation in just 4 of 74 ZZ Cetis ($<$6\,per\,cent) in this manuscript.

This decrease in detection among ZZ Cetis with TESS can be understood in two ways. First, we are generally limited to a Nyquist frequency of 4166.67\,$\mu$Hz (1/240 s$^{-1}$) with our 2-min-cadence data, so we are biased against the shortest-period ZZ Cetis that more commonly exhibit clear rotational splittings \citep{2017ApJ...841L...2H}. Additionally, and likely most significantly, the noise limits (and thus the lowest amplitudes we can significantly detect) from TESS data on our ZZ Cetis are generally an order of magnitude worse than the 27 ZZ Cetis in \citet{2017ApJS..232...23H} --- in that work the mean significant threshold is 0.75\,ppt, whereas the mean value in this work is 8.4\,ppt. We generally have the frequency resolution to detect 0.5-2.0-day rotation periods, but are only able to detect the highest-amplitude modes, and thus miss identifying many rotationally split multiples.

With additional coverage and high-enough cadence to detect shorter-period pulsations, we are optimistic that future observations of white dwarfs with the 20-second cadence will enable us to detect rotationally split modes in far more ZZ Cetis going forward.


\section{Analysis of the sample }
\label{section6}

\begin{figure}
	\includegraphics[width=\linewidth]{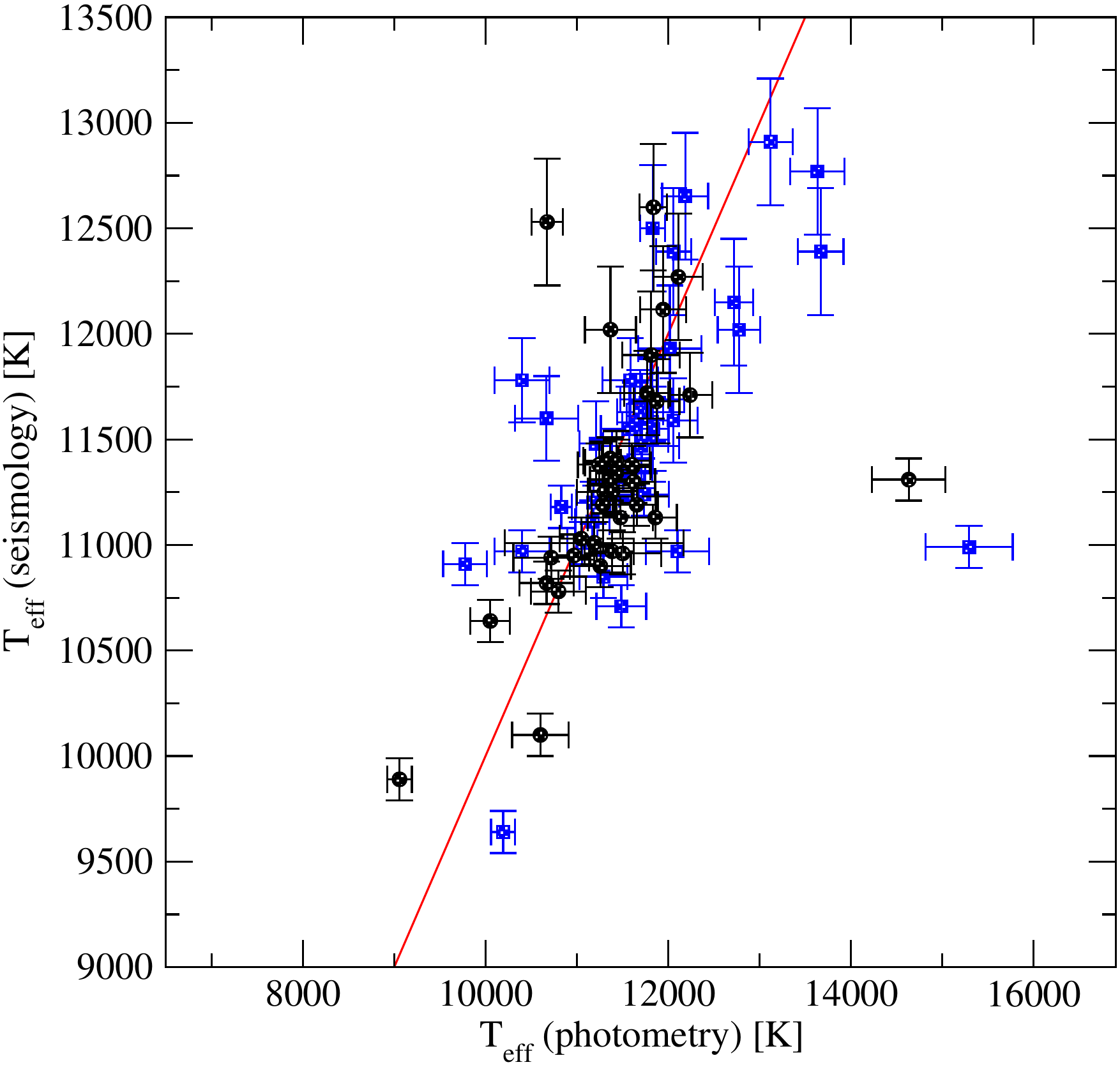}
    \caption{Comparison between the effective temperature obtained from Gaia photometry + parallax (see Table~\ref{tab:list1}) and the asteroseismological fit (see Table~\ref{tab:list1-res}). The red line indicates the 1:1 correspondence. Blue squares correspond to objects with one or two detected periods, and black circle to those with more than two detected periods. }
    \label{tef-comp}
\end{figure}

In this section we analyse the main results of our sample of 75 new bright ZZ Ceti stars reported in this paper, corresponding to the 74 objects observed by TESS and TIC~20976653. In Figure~\ref{tef-comp} we compare the values for the effective temperature from photometry + parallax from Gaia (x-axis) and asteroseismology (y-axis). We consider that the internal uncertainties from the asteroseismological fitting procedure are 100~K, 200~K, and 300~K, for effective temperatures, below $11\, 400$ K, between  $11\, 400$ and $11\, 800$ K, and higher than $11\, 800$~K, respectively. The uncertainties for the photometric effective temperature are taken from Table~\ref{tab:list1}. We depict the objects with one or two detected periods with blue squares, while the objects with more than two detected periods are depicted with black circles.
As can be seen from this figure, the data clusters around the 1:1 correspondence line. The outliers correspond to those objects with photometric mass below 0.45~M$_{\odot}$ and those with photometric effective temperature higher than $14\,000$ K. The Pearson coefficient is $r$ = 0.1154, which indicates a negligible correlation. We do not expect a full correlation since both determinations come from different data sets: the three photometric filters and parallax for the photometric determination, and the detected period spectrum for the seismological determination. 

\begin{figure}
	\includegraphics[width=\linewidth]{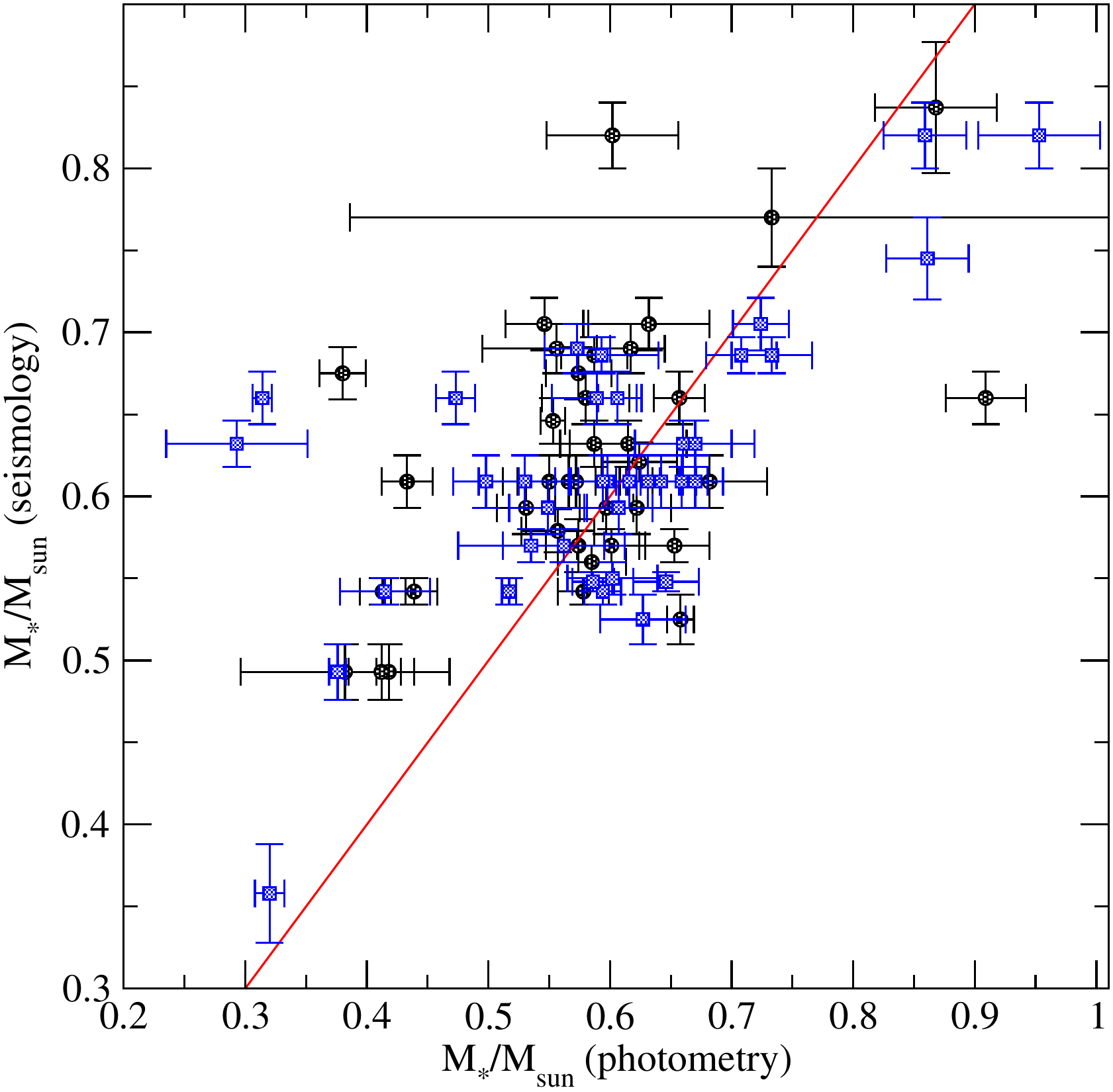}
    \caption{Comparison between the stellar mass obtained from Gaia photometry + parallax (see Table~\ref{tab:list1}) and the asteroseismological fit (see Table~\ref{tab:list1-res}). The red line indicates the 1:1 correspondence. The objects with one or two detected periods are depicted as blue squares, while those with more than two detected periods are depicted with black circles. }
    \label{comp-mass}
\end{figure}

Figure~\ref{comp-mass} shows the comparison between the stellar mass from photometry + parallax (x-axis) from Gaia and the value obtained from our seismological fit. The black circles correspond to the 36 objects with more than two detected periods, while the blue squares depict the 37 objects with one or two detected periods; these blue squares have seismological results that are less reliable since there are so few constraints from pulsation periods. Note that we do not include TIC~345202693 since this object has no reliable photometric parameters due to its main sequence companion. The uncertainties for the seismological mass correspond to internal uncertainties from the fitting procedure. Although the points are around the 1:1 correspondence line, there is a large scatter. The Pearson coefficient is $r$= 0.6020, corresponding to a moderate correlation. 

The larger discrepancies between the photometric and seismological masses appear for the objects with photometric mass below 0.45~M$_{\odot}$. Since our model grid do not consider white dwarf models with stellar masses below 0.493~M$_{\odot}$, we do not expect an agreement between the two determinations. 

\begin{figure}
	\includegraphics[width=\linewidth]{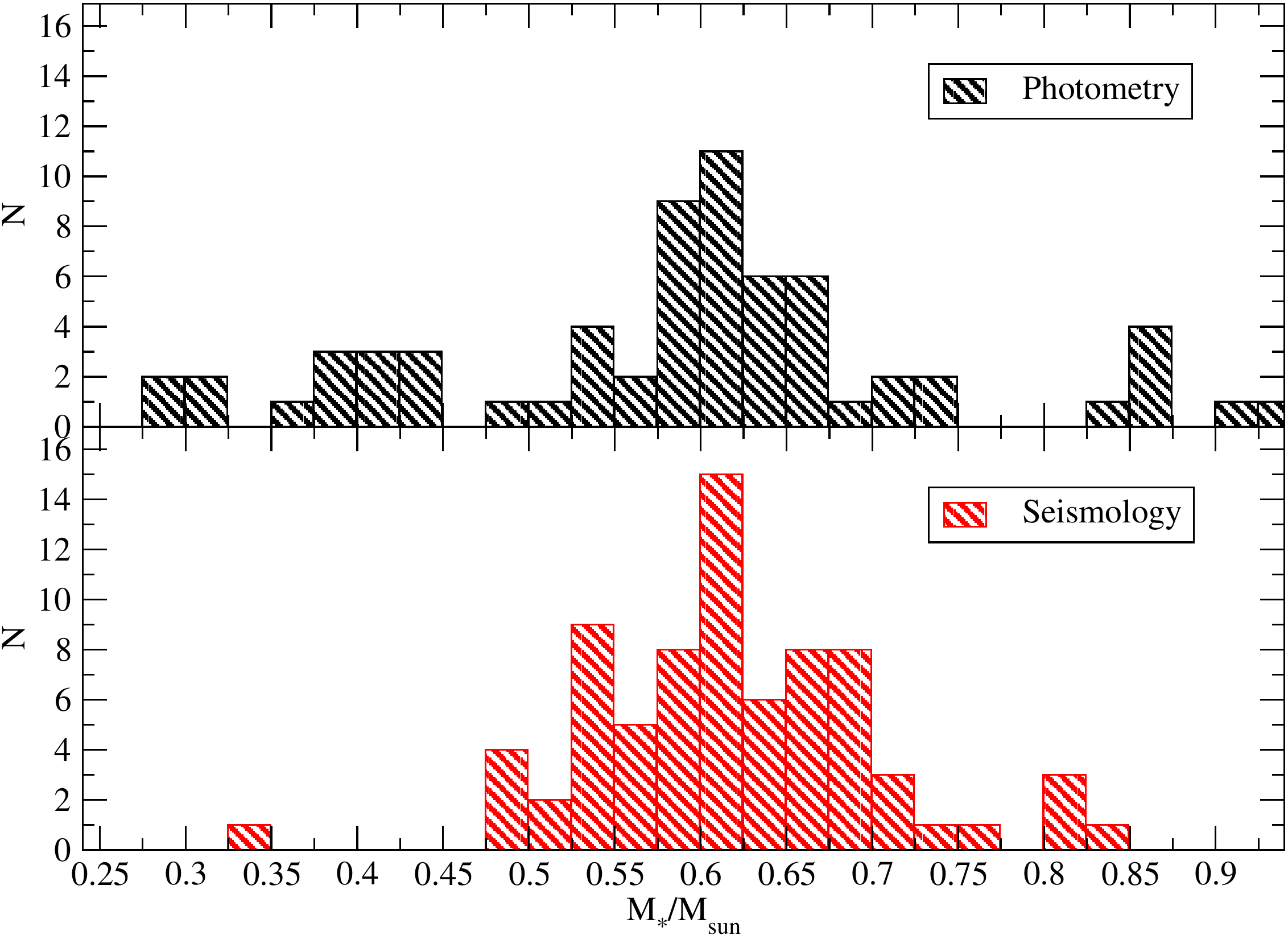}
    \caption{Histograms showing the distribution of the photometric (top) and seismological (bottom) stellar mass. }
    \label{histo-mass}
\end{figure}

The mass distribution for our 74 objects is shown in Figure~\ref{histo-mass}, where we show the histograms for the photometric (top panel) and the seismological (low panel) stellar mass. As expected, the mass distribution from photometry extends further to lower stellar masses, than the one from asteroseismology. In both cases, most of the object show stellar masses between 0.5 and 0.7~M$_{\odot}$.

The mean photometric mass is $\langle M_{\rm phot}\rangle =$ 0.588$\pm$ 0.038 M$_{\odot}$, while for the seismological mass, the value is $\langle M_{\rm seis} \rangle =$ 0.621$\pm$0.015 M$_{\odot}$. Even though both values agree within the uncertainties, the seismological mean mass is $\sim$ 5 per\,cent higher than the photometric value. If we consider only the 61 objects with photometric masses larger than 0.49 M$_{\odot}$, the values are $\langle M_{\rm phot} \rangle =$ 0.631$\pm$ 0.040 M$_{\odot}$, and $\langle M_{\rm seis} \rangle =$ 0.635$\pm$0.015 M$_{\odot}$, with an agreement within 1$\sigma$. Finally, these values for the photometric and seismological mean mass are in agreement with the mean mass of 351 known ZZ Ceti stars shown in Figure \ref{ZZCetis} with coloured symbols, being $\langle M_* \rangle =$ 0.644$\pm$0.034 M$_{\odot}$.

\section{Conclusions}
\label{conclusions}

In this work we present the discovery of 74 new ZZ Ceti stars, based on the data from the TESS mission, from Sector 1 to Sector 39. In addition, we perform follow-up observations for 11 objects from ground-based facilities, i.e., the Konkoly observatory (1.0-m), SOAR telescope (4.1-m), Perkins telescope (1.8-m) and the Pico dos Dias observatory (1.6-m), which in most cases, increased the number of detected periods.  The new ZZ Cetis are much brighter than the average previously known ZZ Ceti, and in this sample range from $13.5<G<17.5$ mag. 
In addition, we detected one additional new ZZ Ceti, TIC~20979953, from ground-based observations showing three pulsation periods. This object has no observations yet with TESS.

We perform a preliminary asteroseismological study of the new sample, and determine their seismological stellar mass, effective temperature and hydrogen envelope mass, among other structural parameters, depending on the number of detected periods. Extensive observations are required to detect a significant number of periods for a more meaningful seismological study, which will in many cases be enabled simply by adding future TESS data, which will improve noise limits to allow us to detect more modes.

We detected rotational splittings from TESS data for four objects, TIC~7675859, TIC~21187072, TIC~343296348 and TIC~394015496. Our derived rotation periods ($0.2-2.2$ days) are roughly compatible with previous estimates of other white dwarf stars. 

The mean stellar mass of our sample from photometry and seismology are $\langle M_{\rm phot}\rangle =$ 0.588$\pm$ 0.038 M$_{\odot}$ and $\langle M_{\rm seis} \rangle =$ 0.621$\pm$0.015 M$_{\odot}$, respectively. Considering the 61 objects with photometric masses above 0.49 M$_{\odot}$, the values are $\langle M_{\rm phot} \rangle =$ 0.631$\pm$ 0.040 M$_{\odot}$, and $\langle M_{\rm seis} \rangle =$ 0.635$\pm$0.015 M$_{\odot}$, respectively. Both values are in agreement with the mean spectroscopic mass of a sample of 351 known ZZ Ceti depicted in Figure \ref{ZZCetis}, $\langle M_* \rangle =$ 0.644$\pm$0.034 M$_{\odot}$.

These 75 new bright ZZ Cetis increase the sample of known pulsating DA white dwarf stars by roughly 20 per\,cent, and our understanding of their interiors will only improve with additional observations from the TESS mission.

\section*{Acknowledgements}

This study was financed in part by the Coordena\c{c}\~ao de Aperfei\c{c}oamento de Pessoal de N\'{\i}vel Superior - Brasil (CAPES) - Finance Code 001, Conselho Nacional de Desenvolvimento Cient\'{\i}fico e Tecnol\'ogico - Brasil (CNPq), and Funda\c{c}\~ao de Amparo \`a Pesquisa do Rio Grande do Sul (FAPERGS) - Brasil. KJB is supported by the National Science Foundation under Award AST-1903828. IP acknowledges support from the UK's Science and Technology Facilities Council (STFC), grant ST/T000406/1. Financial support from the National Science Centre under project No.\,UMO-2017/26/E/ST9/00703 is acknowledged. J.J.H. acknowledges salary and travel support through {\em TESS} Guest Investigator Programs 80NSSC19K0378 and 80NSSC20K0592, and SOAR observational time through NOAO programs 2019B-0125 and 2021B-007.
M.U. acknowledges financial support from CONICYT Doctorado Nacional in the form of grant number No: 21190886 and ESO studentship program.
ZsB acknowledges the financial support of the Lend\"ulet Program of the Hungarian Academy of Sciences, projects No. LP2018-7/2021 and LP2012-31, the KKP-137523 `SeismoLab' \'Elvonal grant of the Hungarian Research, Development and Innovation Office (NKFIH), and the J\'anos Bolyai Research Scholarship of the Hungarian Academy of Sciences.
Based on observations obtained at Las Campanas Observatory under the run code 0KJ21U8U.
Based on observations obtained at the Southern Astrophysical Research (SOAR) telescope under the program allocated by the Chilean Time Allocation Committee (CNTAC), no: CN2020A-87, CN2020B-74 and CN2021A-52. 
Based on observations at the Southern Astrophysical Research (SOAR) telescope, which is a joint project of MCTIC–Brazil, NOAO–US, the University of North Carolina at Chapel Hill (UNC), and Michigan State University (MSU), and processed using the IRAF package, developed by the Association of Universities for Research in Astronomy, Inc.,
 under a cooperative agreement with the US National Science Foundation.This paper includes data collected with the TESS mission, obtained from the MAST data archive at the Space Telescope Science Institute (STScI). Funding for the TESS mission is provided by the NASA Explorer Program. This work has made use of data from the European Space Agency (ESA) mission Gaia (\url{https://www.cosmos.esa.int/gaia}), processed by the Gaia Data Processing and Analysis Consortium (DPAC, \url{https://www.cosmos.esa.int/web/gaia/dpac/consortium}). Funding for the DPAC has been provided by national institutions, in particular the institutions participating in the Gaia Multilateral Agreement. 
This research has made use of NASA's Astrophysics Data System Bibliographic Services, and the
SIMBAD database, operated at CDS, Strasbourg, France,

\section*{Data Availability}

Data from TESS is available at the MAST archive \url{https://mast.stsci.edu/search/hst/ui/$#$/}. Ground based data will be shared on reasonable request to the corresponding author.



\bibliographystyle{mnras}
\bibliography{references} 








\bsp	
\label{lastpage}
\end{document}